\documentclass{article}

\usepackage{amsthm,amsmath,amssymb}
\usepackage{algorithm}
\usepackage{algorithmic}

\usepackage{graphicx}

\begin{document} 

\title{How stochastic synchrony could work in cerebellar Purkinje cells}

\author{Sergio Verduzco-Flores \\
Computational Cognitive Neuroscience Laboratory, \\
Department of Psychology and Neuroscience, \\ 
University of Colorado Boulder, Boulder CO, USA \\
sergio.verduzco@gmail.com
}

\date{ }

\maketitle

\begin{abstract} 
Simple spike synchrony between Purkinje cells projecting to a common
neuron in the deep cerebellar nucleus is emerging as an important
factor in the encoding of output information from cerebellar cortex.
Stochastic synchronization is a viable mechanism through which this
synchrony could be generated, but it has received scarce attention,
perhaps because the presence of feedforward inhibition in the input
to Purkinje cells makes insights difficult.
This paper presents a method to account for feedforward inhibition
so the usual mathematical approaches to stochastic synchronization
can be applied. Three concepts (input correlation, heterogeneity, 
and PRC shape) are then introduced to facilitate an intuitive understanding
of how different factors can affect synchronization in Purkinje cells.
This is followed by a discussion of how stochastic synchrony could
play a role in the cerebellar response under different assumptions.
\end{abstract}



\section{Introduction}
The cerebellum has a striking and relatively clear anatomical organization, 
which has brought hope that it could be the first brain system whose function
could be understood in terms of its structure \cite{ito_cerebellar_2006}. 
There is agreement that the cerebellum may play a role in a variety of
cognitive functions, in addition to its involvement in motor control
\cite{koziol_consensus_2014}.

\begin{figure}
	\includegraphics[width=4in]{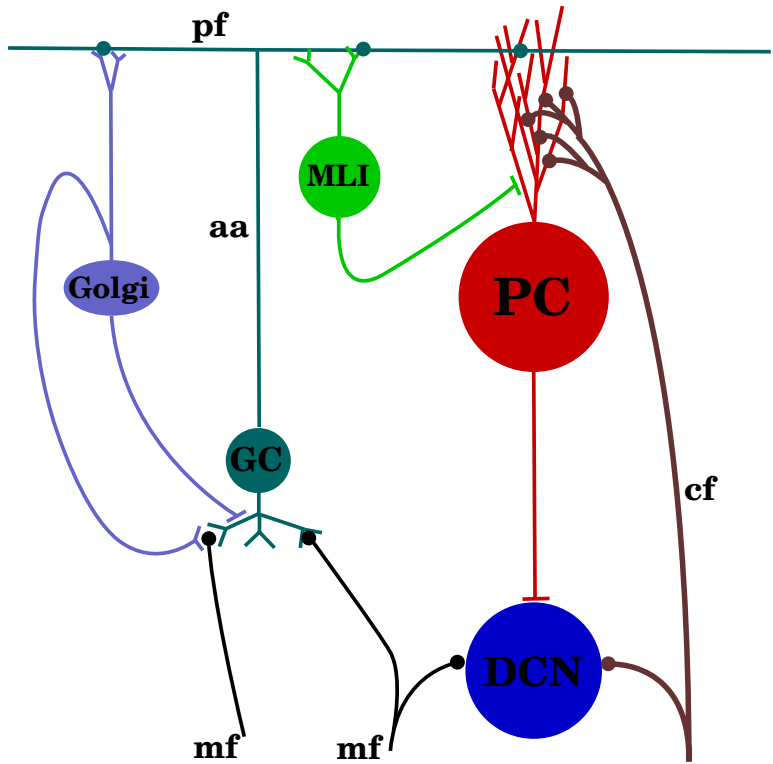}
\caption{
	{\bf Basic connection scheme of the cerebellum. }
Granule cells (GC) receive afferent and efferent information from the mossy fibers
(MF), and convey that information through their parallel fibers (PF). The PFs 
excite both Purkinje Cells (PC) and molecular layer interneurons (MLI); in turn
the MLIs send axons to the PCs. Purkinje cells constitute the only output of the 
cerebellar cortex, and they send axons that form GABAergic inhibitory 
connections on the cells of the deep cerebellar nuclei (DCN). Neurons in
the inferior olivary nucleus (IO) send axons known as 
climbing fibers (CF) which form thousands of synapses on PCs. Each PC receives 
excitation from a single CF. An action potential in a CF reliably causes an action 
potential in
the PCs it innervates; these action potentials are known as complex spikes, and are
easily distinguishable from simple spikes, which are action potentials tonically
generated by the PC, and modulated by the PFs.
Not illustrated in the figure is the fact that the ascending axons (AA) of the
granule cells can make multiple connections on the dendritic arbor of PCs
\cite{bower_organization_2002}, the orientation of PC dendritic arbors perpendicular
to parallel fibers, or the organization in parasagittal modules 
\cite{ruigrok_ins_2011}.
}
\label{anat}
\end{figure}

Figure \ref{anat} shows the basic anatomical organization of the cerebellar cortex.
See \cite{voogd_anatomy_1998} or \cite{ito_cerebellar_2006} for reviews. 
In order to designate specific types of neurons and axons in the cerebellum the 
abbreviations of figure \ref{anat} will be used. Synapses from a source neuron/axon
type towards a target neuron type will be denoted  by the
abbreviation of the source and target connected by a dash; e.g. PF-PC denotes
the synapse between parallel fibers and Purkinje cells.

Perhaps the most influential set of ideas regarding how the cerebellum works is the 
Marr/Albus model \cite{marr_theory_1969,albus_theory_1971}, which has led to a 
variety of models in which Purkinje cells
act like a perceptron whose learning signal comes from the climbing fibers.
(e.g. \cite{fujita_adaptive_1982,brunel_optimal_2004,dean_cerebellar_2010,
walter_advantages_2009,steuber_cerebellar_2007,yamazaki_computational_2012,
Schweighofer_role_1998}).
The discovery of conjunctive Long-Term Depression (LTD) in the PF-PC synapses 
\cite{ito_climbing_1982,ito_synaptic_1993,hirano_long-term_2013}
has added plausibility 
to these models. Over time, however, it has become increasingly clear that conjunctive 
LTD alone may not explain learning in the cerebellum. 

Several studies suggest the incompleteness of conjunctive LTD to explain cerebellar 
learning.  First, it has been shown that cerebellar motor learning can take place in the
absence of PF-PC LTD \cite{welsh_normal_2005,schonewille_reevaluating_2011}.  Second, the
correlation of Purkinje cell firing and muscle EMG can show both positive or negative
correlations, with positive correlations being the more prevalent 
\cite{miller_role_2002,holdefer_dynamic_2009}. 
If the role of Purkinje cells was to gate motor commands
through just inhibition, negative correlations should be the most common.  Third, synaptic
inhibition of Purkinje cells, whose complex spike-elicited plasticity acts to counteract
PF-PC conjunctive LTD, seems to play a role in motor learning \cite{wulff_synaptic_2009}, 
which is ignored by models that rely exclusively on PF-PC LTD.
Moreover, other studies suggest that the timing of Purkinje cells' spikes is important,
not only their firing rate. Tottering mutant (tg) mice have virtually
the same firing rate as that of wild types during spontaneous activity and in response to
optokinetic stimulation; nevertheless, tg mutants show abnormal compensatory eye movements
and severe ataxia \cite{Hoebeek_increased_2005}.

When explaining how the timing of PC simple spikes affect their DCN targets, 
and how different types of CF-mediated plasticity affect cerebellar output, 
it may be important to
pay attention to synchrony among Purkinje cells innervating the same DCN cell.
This syncrony can modulate the response of the target DCN cell 
\cite{gauck_control_2000,jaeger_mini-review:_2011,
steuber_determinants_2011,person_purkinje_2012}.
It has been observed that there exists simple-spike synchrony among PCs
separated by several hundred micrometers, and that this synchrony seems
to depend on afferent input \cite{mackay_integrative_1976,ebner_correlation_1981,
heck_-beam_2007,wise_mechanisms_2010,bosman_encoding_2010}.
This synchrony does not seem to be fully explained by 
firing rate co-modulation or PC recurrent collaterals. Firing rate modulation
may be insufficient in this case, because there are cases where the modulation
in synchrony is unrelated to the modulation in firing rate 
\cite{ebner_correlation_1981,heck_-beam_2007}.
Purkinje cell recurrent collaterals tend to
generate oscillations whose coherence decays with distance 
\cite{de_solages_high-frequency_2008}, which is inconsistent with the distances 
across which synchrony is found; moreover, it is unclear how sensory inputs could
modify the functional coupling of Purkinje cells \cite{wise_mechanisms_2010} 
if this coupling depended on a fast oscillatory regime.
It is thus appropriate to study stochastic synchronization as a candidate 
mechanism to explain how Purkinje cells can activate synchronously.

The phenomenon of stochastic synchrony happens when several uncoupled
oscillators synchronize their phases when receiving correlated inputs
\cite{pikovsky_synchronization:_2003,teramae_robustness_2004}. The
intuition behind this is that if the oscillators become entrained to the
inputs then they will respond similarly, thus acquiring similar phases. 
An interesting aspect of stochastic synchronization is that
the degree of synchrony can be controlled by the way that the oscillators
respond to inputs, which opens the possibility of its modulation by
plasticity mechanisms. If synchrony plays a role in shaping the 
response of cerebellar cortex, it seems feasible that there are plasticity 
mechanisms capable of creating synchrony.

One possible reason why stochastic synchrony has not been largely considered in the
case of Purkinje cells is the complication arising from the feedforward inhibition
in the parallel fibers. As shown in figure \ref{anat}, PFs stimulate MLIs, which in
turn stimulate PCs. This inhibition has been observed as IPSPs arising shortly after
EPSPs \cite{mittmann_feed-forward_2005}, and seems to be fundamental in understanding 
the response of PCs \cite{de_schutter_using_1999,santamaria_feedforward_2007}.
Although considerable advances have been made in understanding stochastic synchrony 
\cite{pikovsky_synchronization:_2003,teramae_robustness_2004,
nakao_synchrony_2005,nakao_noise-induced_2007,galan_stochastic_2007,
marella_class-ii_2008,arai_phase_2008,abouzeid_type-ii_2009,ly_synchronization_2009,
hata_optimal_2011,bressloff_stochastic_2011,burton_intrinsic_2012,
zhou_impact_2013,kurebayashi_phase_2013}, 
no study has explored how feedforward inhibition affects this process.

Exploring stochastic synchronization of Purkinje cells requires to represent their
activity in terms of their {\it phase}. A neuron that fires periodically 
can be understood as a dynamical system whose
trajectory in phase space follows an asymptotically stable limit cycle. Such a
dynamical system can be described by a single variable called its {\it phase};
the phase describes how far the current state is along the limit cycle trajectory.
Perturbations to the system (such as synaptic inputs in the case of a
neuron) can be described by how they shift the phase of the system when they
are received \cite{winfree_geometry_2001,kuramoto_chemical_1984}. The PRC (Phase
Response Curve or Phase Resetting Curve) of the system plots the shift
in phase that an input produces as function of the system's phase when the
input is received. PRCs are a standard tool when understanding
the behavior of coupled oscillators, and have been extensively used to
describe networks of neurons \cite{schultheiss_phase_2012,smeal_phase-response_2010}.
Also, as expected, PRCs are also a standard tool in analytical studies of stochastic 
synchronization.

This paper presents two main sets of results. The first one is the development
of an {\it equivalent PRC}. If the effect of feedforward excitation coming from the
PFs to the PCs is represented with an excitatory PRC, and the effect of feedforward
inhibition coming to the PCs from the MLIs is represented with an inhibitory
PRC, the equivalent PRC lumps the effect of both excitatory and delayed inhibitory
PRCs so we only have one type of inputs. This allows the insights from the 
theory of stochastic synchronization to be applied when there is 
feedforward inhibition, as is the case of Purkinje cells. The equivalent PRC
is defined in two different ways, and in order to validate these definitions
Monte Carlo simulations are performed to verify that an oscillator using the
equivalent PRC responds similarly to an oscillator with excitatory and delayed
inhibitory inputs.

The second set of results in this paper consists of computational simulations of 
oscillators receiving correlated noise and feedforward excitation and delayed 
feedforward inhibition. These simulations show that, as expected, the results from
the theory of stochastic synchronization can provide valuable insights about
the factors which cause synchrony. In order to explore how stochastic synchrony 
may affect cerebellar output, I focus on 3 factors (input correlation, 
heterogeneity, and PRC shape) that control the level of 
synchronization, and on how they may be affected by CF-mediated plasticity.
The output of the synchronizing oscillators is in turn directed towards an
oscillator representing a DCN cell, which includes a simple mechanism to
produce rebound spikes (spikes caused from powerful depolarizing currents activated
by membrane hyperpolarization; they can raise the firing rate in response to
an increase in synchrony). Through
the use of this model it is shown that synchronization is a subtle response
that can depend on several physiological details not often considered
in the literature. For example, for the parameters used: 
(1) synchrony among Purkinje cells does not have to
be related to their firing rate modulation, and (2) the effect of synchrony could 
complement or oppose the putative effect of LTD in the PF-PC synapse.
The paper ends with a discussion of how some hypotheses regarding
cerebellar function may change when synchrony is taken into account.


\section{Models}

The equivalent PRC mentioned above will be developed in the next three subsections.
This equivalent PRC can only emulate the effects of excitation and delayed inhibition
in an approximate manner, and there are several ways to define it. This paper
presents two different definitions for the equivalent PRC. The first subsection of
this section presents introductory material, and the next two subsections develop the
two different definitions of the equivalent PRC.

\subsection{Reduction to a phase equation, and the stationary phase PDF}
This subsection briefly outlines some basic results from 
\cite{kuramoto_chemical_1984,winfree_geometry_2001} using notation based on 
\cite{nakao_synchrony_2005}. The results show how $N$ dynamical systems
oscillating in a limit cycle and receiving impulsive inputs can be represented
with $N$ phase variables. PRCs, the phase transition function, and the phase
evolution equation are introduced for individual oscillators, which allows to 
find an equation for their stationary phase Probability Density Function (PDF).
If we measure the phase of the oscillator at some random point in time, the phase 
PDF can provide the probability that the sampled phase is in a particular interval.

\noindent Consider $N$ oscillators whose dynamics can be expressed as
\begin{equation}
	\bold{\dot X}_i(t) = \bold{F}(\bold{X}_i(t)) + \bold{I}_i(t)
\end{equation}
for $i= 1,\dots,N$, where the vector $\bold{X}_i(t)$ denotes the state of
oscillator $i$ at time $t$, $\bold{F}$ 
is the function describing the dynamics of each oscillator, and
$\bold{I}_i(t)$ represents external random inputs consisting of impulsive
displacements in phase space. We assume that the oscillators have
an asymptotically stable limit cycle $\bold{X}_0(t)$. The impulsive inputs 
are given by

\begin{equation}
	\bold{I}_i(t) = \sum_{n=1}^\infty \bold{e}_n^i \delta(t - t_n^i)
\end{equation}
where $t_n^i$ represents the time of the $n$-th input to the $i$-th
oscillator, and $\bold{e}_n^i$ provides the direction of the shift in 
phase space caused by the corresponding input, so that at time
$t_n^i$ the oscillator $i$ receives an immediate shift in phase
space from point $\bold{X}_i$ to point $\bold{X}_i + \bold{e}_n^i$. 
Since our oscillator represents a neuron, the value of $\bold{e}_n^i$
should be determined by the synapse that receives the input. Furthermore,
it is assumed that the inputs will not take the system outside the basin
of attraction of $\bold{X}_0(t)$.
The inputs that will be considered in this paper behave as Poisson
random point processes. If an input has a mean firing rate $r$, then
its interimpulse interval $T$ has an exponential distribution:
\begin{equation}
	P(T) = r \text{e}^{-rT}
\end{equation}
where $P(T)$ is the probability density function for $T$.

\noindent We define a phase variable $\theta$ along the limit cycle
so that $\theta(t) = \theta(\bold{X}_0(t))$, $\theta$ has a constant angular
velocity $\omega$, and its range is $[0,1)$. This means that in the
absence of external inputs we'll have:
\begin{equation}
	\dot \theta_i(t) = \omega_i.
\end{equation}
In the first part of the results section we work with systems where all the
oscillators have the same angular frequency $\omega$.
The phase of points not directly on $\bold{X}_0$ is defined through
the use of isochrons, which are the set of points that asymptotically
converge to a particular trajectory $\bold{X}^*_0$ in the periodic orbit. 
If the points on an isochron converge to the trajectory $\bold{X}^*_0$, then
their phase at time $t$ is $\theta(\bold{X}^*_0(t))$.
For all functions in this paper whose arguments include a phase value, 
I assume that this phase value is taken modulo 1; e.g. phases $0$,
$1$ are the same phase, and the same can be said of phases $-0.1$,
$0.9$, and $2.9$.
When an oscillator has phase $\theta$ at time $t$, and an
input shifts its state at that moment by an amount $\bold{e}$ then
this state moves to a new isochron, with the 
consequent shift in phase denoted by $G(\theta,\bold{e})$.
We define the phase transition function as
\begin{equation}
	F(\theta,\bold{e}) = \theta + G(\theta,\bold{e}),
\end{equation}
with its output taken modulo 1.
If the phase of an oscillator at time $t_n$ right before its $n$-th input
is $\theta_n$, then the phase right after the input can be written
as $\phi_n \equiv F(\theta_n,\bold{e}_n) = \theta_n + G(\theta_n,\bold{e}_n)$.
The evolution equation of the phase is an iterative equation describing
the phase of the oscillator at the time when the $(n+1)$-th input arrives:
\begin{equation}
	\theta_{n+1} = \omega T_n + F(\theta_n,\bold{e}_n)
	= \theta_n + \omega T_n + G(\theta_n,\bold{e}_n),
\end{equation}
where $T_n = t_{n+1}-t_n$.
The dynamics of the phase in continuous time are described by the equation:
\begin{equation}
	\dot \theta(t) = \omega + \sum_{n=1}^\infty G(\theta_n,\bold{e}_n)
	                 \delta(t-t_n).
\end{equation}
Given that the input times $t_n$ and the input effects $\bold{e}_n$ are 
random variables, so are the phases $\theta_n$. The Probability Density
Function (PDF) of the phase $\theta$ at time step $n$ is denoted 
$\rho(\theta,n)$. This PDF can be described by the following generalized
Frobenius-Perron equation \cite{lasota_probabilistic_2008}:
\begin{equation}
	\rho(\theta,n+1) = \int_0^1 W(\theta - \phi) \int Q(\bold{e})
	\int_0^1 \delta(\phi - F(\psi,\bold{e}))\rho(\psi,n)
	\ d\psi \ d\bold{e} \ d\phi.
\end{equation}
The term $W(\theta - \phi)$ represents the probability that during the
interimpulse interval $T_n$ the phase changes from $\phi$ to $\theta$.
$Q(\bold{e})$ is the probability density function for $\bold{e}$. 
Intuitively, the two innermost integrals produce the expected phase
after the $n$-th input with $\psi$ being the starting phase; this expected
phase, represented by the integration variable $\phi$ is taken in the
outermost integral in order to calculate the probability that during
the interspike interval the phase transitions from $\phi$ to $\theta$.
The transition kernel $W(\theta)$ can be explicitly obtained in the 
case of Poisson inputs, considering that its arguments are taken to
be modulo 1:
\begin{align}
	W(\theta) &= \frac{1}{\omega} \sum_{j=0}^\infty 
	P \left( \frac{\theta+j}{\omega} \right) =
	\frac{r}{\omega} \sum_{j=0}^\infty \text{e}^{-\frac{r}{\omega}\theta}
	\text{e}^{-\frac{r}{\omega}j} \notag \\
	&= \frac{A \text{e}^{-A \theta}}{1 - \text{e}^{-A}},
	\label{eq:tran_kern}
\end{align}
with $A = r/\omega$.
In the limit of a large number of transitions the PDFs $\rho(\theta,n)$
will reach a stationary state $\rho(\theta)$ obeying:
\begin{equation}
	\rho(\theta) = \int_0^1 W(\theta - \phi) \int Q(\bold{e})
	\int_0^1 \delta(\phi - F(\psi,\bold{e}))\rho(\psi)
	\ d\psi \ d\bold{e} \ d\phi.
	\label{eq:phase_pdf}
\end{equation}
I refer to this equation as the phase PDF equation.

\subsection{The equivalent PRC as a function of the phase PDF}
The first idea to obtain an equivalent PRC, denoted in this subsection 
as $\Delta$, is to take the phase PDF $\rho(\theta)$ produced by the
excitatory and inhibitory inputs, and define $\Delta$ so its phase
PDF matches $\rho(\theta)$. This idea will be developed below.

I start by assuming a single oscillator and a single Poisson
process that produces excitatory inputs. Feedforward inhibition
is modeled by assuming that for each excitatory input at time $t_n$
there will be a corresponding inhibitory input at time $t_n + d$, where
$d$ represents the feedforward delay. All excitatory inputs will produce
a shift in phase space $\bold{e}_{exc}$, whereas inhibitory
inputs produce a shift $\bold{e}_{inh}$. The excitatory and inhibitory
PRCs are defined respectively as:
$\Delta_{exc}(\theta) = G(\theta,\bold{e}_{exc})$,
$\Delta_{inh}(\theta) = G(\theta,\bold{e}_{inh})$, where the function
$G$ maps shifts in phase space to shifts in phase.
An oscillator using instead the equivalent PRC will present a shift
in phase $\Delta(\theta)$ at the times when the excitatory inputs arrive.

I assume a perturbation from the system where the PRC is zero for all
phases and the phase PDF is uniform. Furthermore, the perturbation is
small enough so that the phase transition function $F$ is still invertible. Let 
$\rho(\theta) = 1 + \varepsilon \rho_1(\theta)$, and expand the equivalent PRC as
$\Delta(\theta) = \varepsilon \Delta_1(\theta) + \varepsilon^2 \Delta_2(\theta)
+ O(\varepsilon^3)$ (using big Oh notation), so that as $\varepsilon$ goes to 
zero we recover the
unperturbed system. Before substituting these terms, the phase PDF equation
\ref{eq:phase_pdf} is simplified in 3 steps:
\begin{enumerate}
	\item The middle integral disappears, since there is only one type of
		input, and the PDF $Q$ integrates to one.
	\item We perform the innermost integral using the basic formula
		for performing change of variables with Dirac $\delta$ functions.
		If $g(x)$ is a real function with a root at $x_0$ the formula is 
		$\delta(g(x)) = \frac{\delta(x - x_0)}{|g'(x_0)|}$.
	\item We substitute the transition kernel $W$ for its expression in Eq.
		\ref{eq:tran_kern}.
\end{enumerate}
This yields:
\begin{equation}
	\rho(\theta) = \int_0^1 \frac{A \text{e}^{-A(\theta-\phi)}}{1 - \text{e}^{-A}}
	\frac{\rho(F^{-1}(\phi))}{|F'(F^{-1}(\phi))|} \ d\phi
\end{equation}
Notice that the argument $(\theta - \phi)$ of $W$ is still taken modulo 1, so we need 
separate integrals for the cases when this argument is positive and negative:
\begin{equation}
	\rho(\theta) = \frac{A \text{e}^{-A \theta}}{1 - \text{e}^{-A}} 
	\left[ \int_0^{\theta} \text{e}^{A \phi} 
	\frac{\rho(F^{-1}(\phi))}{|F'(F^{-1}(\phi))|} \ d\phi +
	\text{e}^{-A} \int_\theta^1 \text{e}^{A \phi} 
        \frac{\rho(F^{-1}(\phi))}{|F'(F^{-1}(\phi))|} \ d\phi \right].
	\label{eq:pdf_b}
\end{equation}
We now assume $F(\phi) = \phi + \varepsilon \Delta_1(\phi) + 
\varepsilon^2 \Delta_2(\phi) + O(\varepsilon^3)$. Let
$F^{-1}(\phi) = \phi + \varepsilon \psi_1(\phi) + \varepsilon^2 \psi_2(\phi) + O(\varepsilon^3)$. 
Substituting these two previous expressions in the identity 
$F(F^{-1}(\phi)) = \phi$\ we obtain:
\begin{align*}
	\psi_1 &= -\Delta_1, \\
	\psi_2 &= -\Delta_2 + \Delta_1 \Delta_1', \\
	F^{-1}(\phi) &= \phi - \varepsilon \Delta_1 + \varepsilon^2
	(\Delta_1 \Delta_1' - \Delta_2) + O(\varepsilon^3),
\end{align*}
where the argument has been omitted in some functions for brevity of notation.
Using this expression for $F^{-1}$ we find:
\begin{equation*}
	\frac{\rho(F^{-1}(\phi))}{|F'(F^{-1}(\phi))|} = 
	\frac{1 + \varepsilon \rho_1 - \varepsilon^2 
	\Delta_1 \rho_1' + O(\varepsilon^3)}
        {1 + \varepsilon \Delta_1' -
	\varepsilon^2(\Delta_2' - \Delta_1 \Delta_1'')
	+ O(\varepsilon^3)},
\end{equation*}
where we use the fact that since $F$ is invertible, $F'>0$. Using long
division to eliminate the quotient and substituting into (\ref{eq:pdf_b})
yields:
\begin{equation}
 \begin{split}
	 1 + \varepsilon \rho_1(\theta) = 
         \frac{A \text{e}^{-A \theta}}{1 - \text{e}^{-A}} 
         \bigg( & \int_0^{\theta} \text{e}^{A \phi} \left[ 
	 1 + \varepsilon(\rho_1 - \Delta_1') - \varepsilon^2
         (\Delta_1 \rho_1' - \Delta_1 \Delta_1'' + \Delta_1' \rho_1 
         - (\Delta_1')^2 + \Delta_2') \right] \ d\phi \quad + \\
	\text{e}^{-A} & \int_\theta^1 \text{e}^{A \phi} \left[ 
	 1 + \varepsilon(\rho_1 - \Delta_1') - \varepsilon^2
         (\Delta_1 \rho_1' - \Delta_1 \Delta_1'' + \Delta_1' \rho_1 
         - (\Delta_1')^2 + \Delta_2') \right] \ d\phi \bigg).
	 \label{eq:pdf_c}
 \end{split}
\end{equation}
The terms in this equation can be grouped according to the power of
$\varepsilon$ that they contain. The zeroth-order terms yield an identity
corresponding to the unperturbed case. The equation corresponding to
the first power of $\varepsilon$ is:
\begin{equation}
	\rho_1(\theta) = 
         \frac{A \text{e}^{-A \theta}}{1 - \text{e}^{-A}} 
	 \left( \int_0^{\theta} \text{e}^{A \phi} 
	 (\rho_1 - \Delta_1') \ d\phi \quad +
	\text{e}^{-A} \int_\theta^1 \text{e}^{A \phi}
	 (\rho_1 - \Delta_1') \ d\phi \right).
	 \label{eq:pdf_d}
\end{equation}
This equation can be differentiated with respect to $\theta$, and in
the resulting equation we can solve for $\Delta_1'$ to obtain:
\begin{equation*}
	\Delta_1'(\theta) = -\frac{1}{A} \rho_1'(\theta).
\end{equation*}
Integrating this provides an expression to evaluate $\Delta_1$:
\begin{equation}
	\Delta_1(\theta) = \frac{1}{A} (C_1 - \rho_1(\theta)),  
	\label{eq:delta1}
\end{equation}
where $C_1$ is an integration constant.
Notice that equation \ref{eq:pdf_d} only provides constraints for the 
derivative of $\Delta_1$, so it can't be used to determine
the value of $C_1$, reflecting the fact that oscillators with 
different frequencies can have the same stationary phase PDF. 
Since the quantity $\varepsilon C_1/A$ becomes a constant term in
$\Delta$, it adds an extra amount of advance or
retardation to the phase whenever an input is received, and we can adjust
its value so that the mean firing rate of the oscillator with two PRCs
matches that of the oscillator with the equivalent PRC.

The equation corresponding to $\varepsilon^2$ in equation \ref{eq:pdf_c} is:
\begin{equation}
	\int_0^{\theta} \text{e}^{A \phi} 
         [\Delta_1 (\rho_1' - \Delta_1'') + \Delta_1' 
	 (\rho_1 - \Delta_1') + \Delta_2'] \ d\phi = 
	 - \text{e}^{-A} \int_\theta^1 \text{e}^{A \phi}
         [\Delta_1 (\rho_1' - \Delta_1'') + \Delta_1' 
	 (\rho_1 - \Delta_1') + \Delta_2'] \ d\phi. 
	 \label{eq:pdf_e}
 \end{equation}
As in the previous case we can differentiate with respect to $\theta$
and solve for $\Delta_2'$, which gives:
\begin{equation}
	\Delta_2' = \frac{1}{A} \left[
	\rho_1 \rho_1' - C_1 \rho_1' - \frac{1}{A}(C_1 \rho_1''
        + \rho_1 \rho_1'') + \rho_1 \rho_1' + \frac{1}{A}(\rho_1')^2 \right].
	\label{eq:delta2p}
\end{equation}
We can use the integration by parts formula 
$\int \rho_1 \rho_1'' = \rho_1 \rho_1' - \int (\rho_1')^2$ 
to find the antiderivative of this expression, which is:
\begin{equation}
	\Delta_2 = \frac{1}{A} \left[ 
	\rho_1 (\rho_1 + \frac{\rho_1'}{A} - C_1 )
	- \frac{C_1}{A} \rho_1' + C_2 \right],
	\label{eq:delta2a}
\end{equation}
where $C_2$ is an integration constant.
Alternatively, we can avoid having another integration constant by
finding the definite integral of equation \ref{eq:delta2p} from
$0$ to $\theta$, obtaining
\begin{equation}
	\Delta_2 = \frac{1}{A} \left[ 
	\rho_1 \left(\rho_1 + \frac{\rho_1'}{A} - C_1 \right) -
	\rho_1(0) \left(\rho_1(0) + \frac{\rho_1'(0)}{A} - C_1 \right)
	- \frac{C_1}{A} (\rho_1' - \rho_1'(0)) \right].
	\label{eq:delta2b}
\end{equation}

Notice that these results are consistent with those of \cite{arai_phase_2008},
where a perturbative expansion is used to go from the PRC to the phase PDF.

Monte Carlo simulations were performed in order to verify that an
oscillator using the $\Delta$ approximation from the formulas above would respond
similarly to an oscillator using $\Delta_{exc}$ and $\Delta_{inh}$ when
provided with the same input, which consisted of a Poisson spike train
with a frequency $r = 600$ Hz. This high rate comes from the assumption that
the input comes from many homogeneous synapses receiving spike trains
at a lower rate. The case of heterogeneous synapses will be treated
further ahead. It is assumed that when the oscillator's phase transitions 
from 1 to 0 a spike is emitted; phase is not allowed to go from 0 to 1
due to inhibition.
The shapes of $\Delta_{exc}$ and $\Delta_{inh}$ were generated as half
sinusoidals with a single peak centered at the phase $\phi = 0.5$, 
corresponding to a type I PRC \cite{hansel_synchrony_1995}; 
$\Delta_{exc}$ is always
positive, and $\Delta_{inh}$ is always negative. The general shape of
$\Delta_{exc}$ is justified by the measurements that have been taken
of Purkinje cells' PRCs at high frequencies \cite{phoka_new_2010}. The
shape of $\Delta_{inh}$ is justified by noticing that stimulation of
MLIs tends to consistently induce an increase in the latency of the
next spike in Purkinje cells 
\cite{mittmann_feed-forward_2005,mittmann_linking_2007}. The formulas in
this paper can nevertheless be applied when the excitatory and inhibitory
PRCs have different shapes.
The derivation of the formulas for $\Delta_{eq}$ requires that 
the amplitudes $a_{exc}$ of $\Delta_{exc}$ and $a_{inh}$ of $\Delta_{inh}$ 
be moderate. The performance of the formulas gradually deteriorates
with larger amplitudes. The values of $a_{exc}$ and $a_{inh}$ were
chosen to be near the limit where the agreement obtained from using
$\Delta_{eq}$ is still acceptable.  The feedforward delay $d$ is taken 
to be 5 ms, which is somewhat larger than usual estimations of 1 or 2 ms. 
This long delay is used to test the limitations of the approaches used
here to obtain $\Delta_{eq}$, particularly the one to be presented in
the next subsection.

\begin{figure}
	\includegraphics[width=4in]{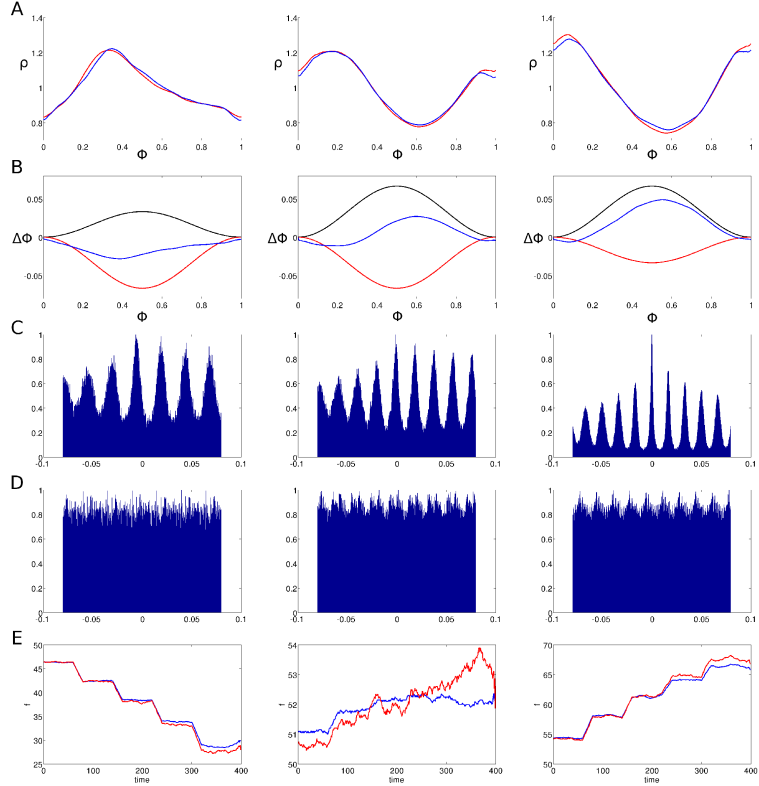}
\caption{
{\bf Comparison of oscillators with 2 and 1 PRCs when matching
phase PDFs.} Simulations for 3 different 
amplitude combinations of $\Delta_{exc}$ and $\Delta_{inh}$. 
For all rows, figures on the
left correspond to $a_{exc} = 1/30, \ a_{inh} = 2/30$. Figures on the
middle correspond to $a_{exc} = 2/30, \ a_{inh} = 2/30$, and Figures on the
right correspond to $a_{exc} = 2/30, \ a_{inh} = 1/30$. The input rate for
all simulations in A-D was $r = 600$ Hz.
A: stationary phase PDFs for oscillators with feedforward inhibition (red) and 
with their equivalent PRC (blue). B: PRCs used in the simulation. $\Delta_{exc}$
(black), $\Delta_{inh}$ (red), $\Delta_{eq}$ (blue). C: Cross-correlograms of the
oscillators' output spike trains with 1 and 2 PRCs. The vertical axis corresponds
to normalized spike count per time bin, and the horizontal axis to time shift.
D: Cross-correlograms of the
output spike train of the oscillator with 2 PRCs and a periodic spike train with 
the same mean frequency. E: Firing rate response of the oscillators 
with 2 (red) and 1 (blue) PRCs to five different levels of input rates. Input 
rates range from 240 Hz to 1200 Hz in constant increments of 240 Hz.
}
\label{result1_1}
\end{figure}

Figure \ref{result1_1} presents the results of substituting the two PRCs
by the equivalent PRC. Panel A shows that there is a good match
between the stationary phase PDFs, as would be expected since the
formulas for $\Delta_{eq}$ were derived with this result in mind. The
PDF curves come from 400 seconds of simulation, which permitted on average
240 000 sample points. 

It seems reasonable that the output spikes of the oscillators with 1 and
2 PRCs would come at similar times, since they have similar rates and 
phase PDFs. This is what is shown in panel C of figure \ref{result1_1},
which shows the cross-correlogram of the output spike trains for both
oscillators. For comparison purposes a similar cross-correlogram
was produced in panel D, between the spike train of the oscillator
with 2 PRCs and a spike train with the same frequency and constant
interspike intervals. It is apparent that the peaks in the top 
cross-correlograms are not just a product of periodicity in the signals.

Finally, I compared the response of both oscillators when the input
firing rates were changed through the simulation. For each simulation
the value of $C_1$ was adjusted up to 2 decimal points so that the 
firing rates of the oscillators would match for an input rate of
600 Hz. As can be seen in panel E, slight inaccuracies in the calculation
of this parameter were amplified for larger firing rates. Moreover,
in the case of balanced excitatory and inhibitory amplitudes the output
firing rate tended to increase for larger input firing rates. This effect
is amplified for larger PRC amplitudes, reflecting the fact that if
the phase shift of $\Delta_{exc}$ is large, the oscillator will spike
before the inhibition arrives.

So far it has been assumed that all inputs of the same type (excitatory
or inhibitory) will produce the same effect on the oscillator. On the other
hand, a real neuron tends to have heterogeneous synapses. One way to represent
this is to have separate excitatory and inhibitory PRCs for each input,
representing different synapses. If
an oscillator has $N_{syn}$ different inputs, we will have PRCs
$\Delta_{exc}^i$, $\Delta_{inh}^i$, for $i = 1,\dots,N_{syn}$. In this case,
for each excitatory/inhibitory pair we may create an equivalent PRC
$\Delta_{eq}^i$. One approach to create $\Delta_{eq}^i$ is to consider the
stationary phase PDF $\rho^i$ and output firing rate that would be produced 
if the inputs with $\Delta_{exc}^i$, $\Delta_{inh}^i$ were considered in 
isolation. The formulas above could then be used to create $\Delta_{eq}^i$.
Instead of doing this, in the next subsection I develop a different way of
obtaining $\Delta_{eq}^i$, based on a more direct calculation of the 
phase-shifting effects produced by feedforward inhibition.

\subsection{Obtaining an equivalent PRC using the expected inhibition}

The method presented above to obtain $\Delta$ is based on creating an
oscillator with a single PRC that has the same phase PDF as the one with
two PRCs. A different idea is as follows. Assume that at time $t$
an excitatory input is received in the oscillator $O_{ei}$ with excitatory and
inhibitory PRCs, and the corresponding inhibitory input is received at time 
$t + d$. We could create an oscillator $O_{eq}$ with an equivalent PRC 
that would apply a phase shift at $t$, and that shift would be such 
that the phase of $O_{eq}$ at time $t + d$ equals the phase of $O_{ei}$
right after the inhibitory input arrives.

It should be clear that this approach can only work on the average. In the
period between $t$ and $t+d$ there may be several inputs shifting the phase,
and the magnitude of the inhibitory shift at time $t+d$ depends on what the 
phase is in that moment. We can then attempt to create a $\Delta_{eq}$
equivalent PRC that on average will lead to being in the same phase as $O_{ei}$ at 
time $t+d$. Such a PRC is constructed below in five stages, each one 
culminating with a version of the equivalent PRC 
that is only appropriate for a restricted set of scenarios. The fifth
and most general version can be used in 
the case of heterogeneous inputs and moderate feedforward delays.
It should be remembered that all phases are interpreted to be modulo 1.

Let us first consider the case of an oscillator with a single Poisson
input and feedforward inhibition. If an excitatory input arrives at
time $t_n$ when the phase is $\phi_n$, then the phase will immediately
experience a shift $\Delta_{exc}(\phi_n)$. At time $t_n + d$ the phase
will be $(\phi_n + \Delta_{exc}(\phi_n) + \omega d)$, where $\omega$ is the
angular frequency of the oscillator. At that moment the inhibitory input
will arrive, causing a phase shift 
$\Delta_{inh}(\phi_n + \Delta_{exc}(\phi_n) + \omega d)$. A simple way
to define $\Delta_{eq}$ for this case is:
\begin{equation}
	\Delta_{eq}(\phi) = \Delta_{exc}(\phi) +
        \Delta_{inh}(\phi+ \Delta_{exc}(\phi) + \omega d).
	\label{eq:eqprc_1}
\end{equation}
This constitutes the first of our for versions for the equivalent PRC.
One difficulty that quickly becomes apparent with it, is that in the time
interval between $t_n$ and $t_n + d$ there will usually be other inputs
arriving at the oscillator, so that the phase when the inhibitory input
arrives will generally not be $(\phi+ \Delta_{exc}(\phi) + \omega d)$.
Indeed, this approach only produces reasonable results when 
inputs are unlikely to arrive between $t_n$ and $t_n+d$, which could happen
when the value of $d$ is very small.

One way to improve our equivalent PRC is to substitute 
$\Delta_{inh}(\phi+ \Delta_{exc}(\phi) + \omega d)$ by the expected value
of the inhibitory shift given the phase when the excitatory shift
happened. Let $\theta(t)$ be the function that gives the phase of the
oscillator at time $t$. Assume that an excitatory input arrives at 
time $t_0$, when the phase is $\phi_0$, meaning $\phi_0 = \theta(t_0)$.
Furthermore, assume that between the times $t_0$ and $t_0+d$ there
arrive $k_e$ excitatory inputs at the times $t_j^e$,
$j = 1,\dots,k_e$; and $k_i$ inhibitory inputs at the times 
$t_m^i$, $m = 1,\dots,k_i$.
For these particular initial phase and inputs define the phase 
deviation as:
\begin{equation}
	D = \sum_{j=1}^{k_e} \Delta_{exc}(\theta(t_j^e)) +
            \sum_{m=1}^{k_i} \Delta_{inh}(\theta(t_m^i)).
	    \label{eq:ph_dev}
\end{equation}
$D$ is a random variable that tells us how much the phase will
change due to inputs during the time time interval
between $t_0$ and $t_0+d$. 
For notational convenience let's define 
$a \equiv \phi + \Delta_{exc}(\phi)$, and
$b \equiv  \phi + \Delta_{exc}(\phi) + \omega d$.
Our goal is to calculate the expected value of 
$\Delta_{inh}(\phi + \Delta_{exc}(\phi) + \omega d + D)$, which is denoted by 
$E(\Delta_{inh}(b + D)|\phi)$. 
This notation indicates the expected value of the inhibitory shift
given that the excitatory shift happened when the phase was $\phi$.
The equivalent PRC can then be defined as:
\begin{equation}
	\Delta_{eq}(\phi) = \Delta_{exc}(\phi) +
        E(\Delta_{inh}(b + D)|\phi).
	\label{eq:eqprc_2}
\end{equation}
This is the second version of the equivalent PRC in this subsection. The 
following paragraphs deal with finding a practical way to calculate
$E(\Delta_{inh}(b + D)|\phi)$, culminating with equation 
\ref{eq:exp_inh_rt}.

To calculate $E(\Delta_{inh}(b + D)|\phi)$, we can start by calculating
this expected value when we know exactly how many excitatory and
inhibitory inputs arrived during the delay period. Assume that the
inputs are independent Poisson point processes, with rate $r_e$ for
the excitatory ones, and rate $r_i$ for the inhibitory ones. Using the
PDF for the Poisson distribution we can obtain:
\begin{equation}
	E(\Delta_{inh}(b + D)|\phi) = 
	\sum_{k_e=0}^\infty \sum_{k_i=0}^\infty
	\left[ \frac{(r_e d)^{k_e}}{k_e!} \text{e}^{-r_e d} \right]
	\left[ \frac{(r_i d)^{k_i}}{k_i!} \text{e}^{-r_i d} \right]
	E(\Delta_{inh}(b+D)|\phi,k_e,k_i),
	\label{exp_inh1}
\end{equation}
where $E(\Delta_{inh}(b+D)|\phi,k_e,k_i)$ is the expected value of the
inhibitory shift given that there were $k_e$ excitatory and $k_i$ inhibitory
inputs during the delay period, making no assumptions about the order in
which they arrived. The assumption of independence between excitatory and
inhibitory inputs is based on the fact that we are restricted to a time
interval of length $d$, during which none of the inhibitory inputs is the
result of feedforward inhibition from one of the excitatory inputs. Notice
that the first two factors decay exponentially, so in practice it is only 
necessary to use a moderate number of terms.

The strategy to obtain $E(\Delta_{inh}(b+D)|\phi,k_e,k_i)$ is to first find
the PDF of $D$, so we can then find the expected value through integration.
This calculation can become involved, so I will first focus on the simpler 
case when there is only a single excitatory input and no inhibitory inputs 
during the delay interval. Under these circumstances, if the 
initial excitatory stimulus arrived at phase $\phi$, the PDF of $D$ is 
denoted by $p(D|\phi,k_e=1,k_i=0)$. What follows is some formal 
reasoning to arrive at an expression for 
$E(\Delta_{inh}(b + D)|a,k_e=1,k_i=0)$. The reader may just go directly to
equation \ref{eq:exp_inh_10}, which is intuitive enough.

Let $\mathcal{B}$ denote the Borel sets  in the interval 
$[0,a_{exc}]$, where $a_{exc}$ is the largest value on the range of
$\Delta_{exc}$, and define
a function $I^e:\mathcal{B} \rightarrow \mathcal{B}$ that maps each set
$A \in \mathcal{B}$ to its preimage under $\Delta_{exc}$.
Given the fact that there was only a single input 
coming from the Poisson process in the phase interval $[a,b]$,
I make the assumption that the input could have arrived with
equal probability at any phase between $a$ and $b$. This implies that
for an interval $H$ in $[0,a_{exc}]$ the probability of $D \in H$ 
is given by $\lambda(I^e(H)\cap [a,b])/\lambda([a,b])$, where 
$\lambda$ is the standard Lebesgue measure for the real numbers.
Notice that $\lambda([a,b]) = \omega d$. If we define a measure
$\mu(H) = \lambda(I^e(H)\cap [a,b])/\omega d$, then the PDF of $D$ will be 
Radon-Nikodym derivative of $\mu$ with respect to $\lambda$. 
A practical way to calculate this PDF starts by partitioning the interval
$[0,a_{exc}]$ into subintervals were $\Delta_{exc}$ is invertible or 
constant, which should be possible for any reasonable PRC. 
If $\Delta_{exc}$ is invertible in the interval $[x,y]$ then 
\begin{equation*}
	\mu([x,y]) = \lambda(I^e([x,y]\cap [a,b]))/\omega d = 
	\frac{|\Delta_{exc}^{-1}(y) -\Delta_{exc}^{-1}(x)|}{\omega d}
	= \frac{1}{\omega d} 
	  \int_{[\Delta_{exc}^{-1}(y),\Delta_{exc}^{-1}(x)]} d \lambda.
\end{equation*}
The change of variables formula shows that
\begin{align*}
        \int_{[\Delta_{exc}^{-1}(y),\Delta_{exc}^{-1}(x)]} d \lambda & =
	\int_x^y \left\lvert \frac{d}{ds} \Delta_{exc}^{-1}(s) \right\rvert \ ds = 
	\int_x^y \frac{1}{|\Delta_{exc}'(\Delta_{exc}^{-1}(s))|} \ ds \\
	& = \left\lvert \int_x^y [\Delta_{exc}'(\Delta_{exc}^{-1}(s))]^{-1} \ ds
	\right\rvert ,
\end{align*}
where the last equation uses the fact that $\Delta_{exc}$ is invertible, so
$\Delta_{exc}'$ doesn't change sign in this interval.

Assume without loss of generality that $\Delta_{exc}' > 0$ in $[x,y]$.
Since 
\begin{equation*}
P(D \in [x,y]|\phi,k_e=1,k_i=0) =\frac{1}{\omega d} 
\int_x^y [\Delta_{exc}'(\Delta_{exc}^{-1}(s))]^{-1} \ ds,
\end{equation*}
then $[\Delta_{exc}'(\Delta_{exc}^{-1}(s))]^{-1}/\omega d$ is the PDF of
$D$ in the interval where $\Delta_{exc}$ is invertible. 
In other intervals the sign of $\Delta_{exc}'$
may be negative, in which case the PDF reverses its sign. If we have
an interval where $\Delta_{exc}$ is equal to a constant $c$, then
$P(D = c|\phi,k_e=1,k_i=0) = \mu(c)$, and finding the expected value 
of the inhibition is trivial. For intervals where $\Delta_{eq}$ is not
constant but is invertible, we have
\begin{equation*}
	E(\Delta_{inh}(b + D)|a,k_e=1,k_i=0) =
	\frac{1}{\omega d} \int_{\Delta_{exc}(a)}^{\Delta_{exc}(b)}
        [\Delta_{exc}'(\Delta_{exc}^{-1}(s))]^{-1} 
	\Delta_{inh}(b + s) \ ds.
\end{equation*}
Using a change of variables this becomes the more intuitive formula
\begin{equation}
	E(\Delta_{inh}(b + D)|a,k_e=1,k_i=0) =
	\frac{1}{\omega d} \int_a^b
	\Delta_{inh}(b + \Delta_{exc}(s)) \ ds.
	\label{eq:exp_inh_10}
\end{equation}
In a similar manner it can be shown that
\begin{equation*}
	E(\Delta_{inh}(b + D)|a,k_e=0,k_i=1) =
	\frac{1}{\omega d} \int_a^b
	\Delta_{inh}(b + \Delta_{inh}(s)) \ ds.
\end{equation*}
The complexity of these equations increases once we have more than
one input, and once we have both excitatory and inhibitory inputs, 
because the order in which they arrive
is important. In this case the expected value for the inhibition comes
from averaging over all the possible phases when the first and second
stimuli could have arrived, and over the possible orders for the 
arrival of stimuli. 
To illustrate this, let's look at the formula for $k_e =1, k_i= 1$
\begin{align*}
	E(\Delta_{inh}(b + D)|\phi,k_e=1,k_i=1) =
	\frac{1}{2 \omega d} \Bigg[
	& \int_a^b \frac{1}{b - \psi_1}
	\int_{\psi_1 + \Delta_{exc}(\psi_1)}^{b + \Delta_{exc}(\psi_1)}
	\Delta_{inh}(b + \Delta_{exc}(\psi_1) + \Delta_{inh}(\psi_2)) 
	\ d \psi_2 d \psi_1 \\ + 
	& \int_a^b \frac{1}{b - \psi_1}
	\int_{\psi_1 + \Delta_{inh}(\psi_1)}^{b + \Delta_{inh}(\psi_1)}
	\Delta_{inh}(b + \Delta_{inh}(\psi_1) + \Delta_{exc}(\psi_2)) 
        \ d \psi_2 d \psi_1 \Bigg].
\end{align*}
Intuitively, the integration variable $\psi_1$ stands for the phase when the
first input arrived, and $\psi_2$ for the phase when the second input
arrived. The first pair of nested integrals are for the case when the 
excitatory input happened first, and the second ones for the case when the 
inhibitory input was the first to arrive. The innermost integrals obtain 
the average inhibition given that the first stimulus arrived at phase $\psi_1$, 
and in the limit when $\psi_1 \rightarrow b$, they converge to the $\Delta_{inh}$
expression with $\psi_1$ substituted by $b$, and $\psi_2$ substituted by
$b + \Delta_{exc/inh}(b)$.

In order to write the formula for the case with arbitrary values for $k_e$ and
$k_i$ some preliminary definitions are required. Notice that if we have
$k_e$ excitatory and $k_i$ inhibitory inputs, there are $C_{k_e}^{k_e+k_i}$
different input sequences according to whether the $j$-th input was
excitatory or inhibitory. Let us index those sequences and denote them by
$\sigma_i$. In other words, we create $C_{k_e}^{k_e+k_i}$ functions
$\sigma_i:\{1,2,\dots,k_i + k_e\} \rightarrow \{-1,1\}$, defined by:
\begin{equation*}
	\sigma_i(j) = 
	\begin{cases}
1, & \mbox{if the }j \mbox{-th element of the }i \mbox{-th sequence is excitatory}; \\
-1, & \mbox{if the }j \mbox{-th element of the }i \mbox{-th sequence is inhibitory}.
	\end{cases}
\end{equation*}
Now define the function
$\Delta_{mix}:[0,1]\times\{1,\dots,C_{k_e}^{k_e+k_i}\}\times\{1,\dots,k_e+k_i\}
\rightarrow R$ (where $R$ stands for the real numbers) by
\begin{equation*}
	\Delta_{mix}(\phi,i,j) = 
	\begin{cases}
		\Delta_{exc}(\phi), & \mbox{if } \sigma_i(j) = 1; \\
		\Delta_{inh}(\phi), & \mbox{if } \sigma_i(j) = -1.
	\end{cases}
\end{equation*}
The general formula can now be written as:
\begin{align}
	& E(\Delta_{inh}(b + D)|\phi,k_e,k_i) = \notag \\
	& \left(\omega d C_{k_e}^{k_e+k_i} \right)^{-1}
	\sum_{i=1}^{C_{k_e}^{k_e+k_i}} \Bigg[
        \int_a^b d \psi_1 \frac{1}{b-\psi_1}
	\int_{\psi_1 + \Delta_{mix}(\psi_1,i,1)}^{b + \Delta_{mix}(\psi_1,i,1)} 
	d \psi_2 \frac{1}{b + \Delta_{mix}(\psi_1,i,1) -\psi_2} \dots \notag \\
	& \int_{\psi_{(k_e+k_i-1)} + \Delta_{mix}(\psi_{(k_e+k_i-1)},i,k_e+k_i-1)}
	^{b + \sum_{m=1}^{k_e+k_i-1}\Delta_{mix}(\psi_m,i,m)} 
	d \psi_{(k_e+k_i)} \Delta_{inh}
	\big(b + \sum_{j=1}^{k_e+k_i}\Delta_{mix}(\psi_j,i,j)\big)
        \Bigg].
	\label{eq:exp_inh_kk}
\end{align}
For this equation I have used the convention of writing the differential
sign next to its corresponding integration sign.

Although equation \ref{eq:exp_inh_kk} expresses the expected inhibition values
that we want to calculate, its complexity makes it virtually useless for
practical purposes. Fortunately, a simple assumption can simplify this 
expression. Assume that for each input sequence, the inputs happen
at regular time intervals (the time periods between any two inputs are 
equal). It is simple to calculate the expected value of the inhibition for
this case. If we have $K = k_e + k_i$ inputs, define 
$\gamma = \omega d/(K+1)$, and for $i = 1,\dots,C_{k_e}^{k_e+k_i}$ let 
\begin{align*}
\theta_0^i &= a, \\
\theta_1^i &= \theta_0^i + \gamma + \Delta_{mix}(\theta_0^i + \gamma,i,1), \\
\theta_2^i &= \theta_1^i + \gamma + \Delta_{mix}(\theta_1^i + \gamma,i,2), \\
& \ \ \vdots \\
\theta_K^i &= \theta_{K-1}^i + \gamma + \Delta_{mix}(\theta_{K-1}^i + \gamma,i,K). 
\end{align*}
We then have:
\begin{equation}
	E(\Delta_{inh}(b+D)|\phi,k_e,k_i,\mbox{RT}) =
	\frac{1}{C_{k_e}^{k_e+k_i}}
	\sum_{i=1}^{C_{k_e}^{k_e+k_i}}
	\Delta_{inh}(\theta_K^i + \gamma),
	\label{eq:exp_inh_rt}
\end{equation}
where RT stands for ``Regular Times'', meaning that the inputs arrive
at regular time intervals.
For a smooth function $\Delta_{inh}$ and relatively small values of
the feedforward delay $d$ we'll have:
\begin{equation*}
	E(\Delta_{inh}(b+D)|\phi,k_e,k_i,\mbox{RT}) 
	\approx E(\Delta_{inh}(b+D)|\phi,k_e,k_i).
\end{equation*}
Even if we now can obtain a good approximation to the expected phase shift
caused by feedforward inhibition, the equivalent PRC from equation
\ref{eq:eqprc_2} may still not achieve the goal of reaching, on average,
the same phase as the oscillator with two PRCs after the feedforward delay.
To make this explicit, assume that the function $\theta(t)$ provides the
phase of the oscillator with feedforward inhibition at time $t$, just as it
is described for equation \ref{eq:ph_dev}, and let $\theta_{eq}(t)$ give the
phase of an oscillator using the corresponding equivalent PRC from equation
\ref{eq:eqprc_2} when the input times are the same. Using an equivalent PRC 
instead of $\Delta_{exc}$ and $\Delta_{inh}$ causes the phase deviation of 
equation \ref{eq:ph_dev} to become
\begin{equation*}
	D_{eq} = \sum_{j=1}^{k_e} \Delta_{eq}(\theta_{eq}(t_j^e)).
	    \label{eq:eqph_dev}
\end{equation*}
In general, $D \neq D_{eq}$ during the delay period; we can calculate the
expected phase difference that this will cause right after the feedforward
delay. If an initial excitatory input arrives at time $t$ when the phase 
is $\phi$, the expected value of the phase for the oscillator with 
2 PRCs at time $t+d$ right after the feedforward inhibition is
\begin{equation*}
b + E(\Delta_{inh}(b+D)|\phi) + E(D|\phi). 
\end{equation*}
On the other hand, the expected value of the phase for the oscillator with 
1 PRC at time $t+d$ is
\begin{equation*}
	b + E(\Delta_{inh}(b+D)|\phi) + E(D_{eq}|\phi). 
\end{equation*}
Subtracting the previous two expressions gives us the expected value of
the phase difference between the two oscillators at time $t+d$ given that
there was an excitatory input at time $t$ when the phase was $\phi$, denoted
by $\xi(\phi)$:
\begin{equation}
	\xi(\phi) = E(D|\phi) - E(D_{eq}|\phi).
	\label{eq:xi1}
\end{equation}
If we are capable of calculating $E(D|\phi)$ and $E(D_{eq}|\phi)$, then we
can use $\xi(\phi)$ in order to create an equivalent PRC that produces
a smaller value of $\xi(\phi)$. A simple algorithm for doing this is
as follows. Let $M$ be an integer, and $\epsilon$ a small real number.
Define $\Delta_{eq}^{(0)}$ as the equivalent PRC from equation \ref{eq:eqprc_2}. 
\begin{algorithmic}
	\FOR{$i$ = 1 to $M$}
	\STATE $\xi^{(i)}(\phi) = E(D|\phi) - E(D_{eq}^{(i-1)}|\phi)$ 
	\STATE $\Delta_{eq}^{(i)}(\phi) = \Delta_{eq}^{(i-1)}(\phi) 
	       + \epsilon \ \xi^{(i)}(\phi) $
	\ENDFOR
\end{algorithmic}
At each step in this algorithm the functions $\xi^{(i)}$ and
$\Delta_{eq}^{(i)}$ are calculated for all the values of $\phi$, so it is
similar to gradient descent performed for a whole function.
The resulting PRC $\Delta_{eq}^{(M)}$ constitutes the third version of
an equivalent PRC we have obtained, and can already provide very
good results for oscillators with a single input, or with homogeneous
excitatory and inhibitory PRCs.

The fourth version of an equivalent PRC that I'll obtain extends the 
second version to the case when there are heterogeneous inputs. The reason
why the approach taken so far to obtain $\Delta_{eq}$ may fail when we
consider several types of inputs, each with its own excitatory and inhibitory
PRCs, is that when equation \ref{eq:exp_inh_rt} is obtained the phase is 
assumed to advance at a steady rate between inputs (the phase would advance
an amount $\gamma$ between inputs). If there is only one type of
input this is justified, since the oscillator has a constant
angular frequency. When we have different types of inputs we 
consider each one separately, so even if inputs of one type arrive at 
regular intervals in time, the phase will be shifted between consecutive
times by inputs of other types. This will become explicit in the 
following calculation.

Consider an oscillator with feedforward inhibition that receives $N_{syn}$ 
different inputs. We consider that for each input there are two ``synapses,''
one excitatory and one inhibitory, characterized by the PRCs
$\Delta_{exc}^i$, and $\Delta_{inh}^i$ for the $i$-th input. 
We need to obtain $N_{syn}$ equivalent
PRCs, with $\Delta_{eq}^i$ being used to replace $\Delta_{exc}^i$ 
and $\Delta_{inh}^i$. The goal is therefore to obtain a version of
equation \ref{eq:exp_inh_rt} that works for each $i$-th input 
individually. In order to model how the oscillator's phase changes 
between repetitions of the $i$-th input I will use the concept of variable
phase velocity, which will be explained next.

Let's say an oscillator has a non-constant phase PDF $\rho(\phi)$. We can
think that this oscillator's phase has a constant rate of change
$d\phi/dt = \omega$, but its inputs reshape the phase PDF so it becomes
$\rho(\phi)$. Alternatively,
we could image that the oscillator receives no inputs, but instead has
a phase whose velocity $d\phi/dt = \omega(t)$ is changing over time so
that $\rho(\phi)$ is produced. The idea to be introduced here is to use
this oscillator with no inputs and variable phase velocity in order to
model the oscillator with constant phase velocity and random inputs. 

Let $T$ denote the period that we will assign to our oscillator. Assume
that at time $t=0$ the oscillator has phase $0$, and denote the time it will
take to reach phase $\phi \in [0,1]$ by $\tau(\phi)$. It is easy to show 
that when $\tau(\phi) = T \int_0^\phi \rho(\psi) \ d\psi$ the phase PDF of
the oscillator is $\rho(\phi)$. Since $\rho > 0$, $\tau(\phi)$ is a monotone
increasing function, which implies that it will be invertible on $[0,1]$. 
Let $\mathcal{P}$ denote this inverse. The function $\mathcal{P}$ provides
the phase as a function of the time since phase $0$ was crossed. 
We consider the argument of $\mathcal{P}$ to be modulo $T$, so it is always in
the interval $[0,T]$. We are now ready to create a version of equation
\ref{eq:exp_inh_rt} for the case of heterogeneous inputs.

Let $\rho(\phi)$ be the phase of the oscillator with feedforward inhibition
and heterogeneous PRCs. Assume that during the feedforward delay period
the synapses for input $j$ receive $k_e^j$ excitatory inputs, and
$k_i^j$ inhibitory inputs. Let $K^j = k_e^j + k_i^j$,  
$\tau_j = \frac{d}{K^j + 1}$.
We can now set
\begin{align*}
\theta_0^{i,j} &= a, \\
\theta_1^{i,j} &= \mathcal{P}(\tau(\theta_0^{i,j}) + \tau_j) + 
     \Delta_{mix}\big(\mathcal{P}(\tau(\theta_0^{i,j}) + \tau_j),i,1 \big), \\
\theta_2^{i,j} &= \mathcal{P}(\tau(\theta_1^{i,j}) + \tau_j) + 
     \Delta_{mix}\big(\mathcal{P}(\tau(\theta_1^{i,j}) + \tau_j),i,2 \big), \\
& \ \ \vdots \\
\theta_K^{i,j} &= \mathcal{P}(\tau(\theta_{K-1}^{i,j}) + \tau_j) + 
     \Delta_{mix}\big(\mathcal{P}(\tau(\theta_{K-1}^{i,j}) + \tau_j),i,K) \big);
\end{align*}
and also:
\begin{equation*}
	E(\Delta_{inh}^j(b+D)|\phi,k_e^j,k_i^j,\mbox{RT})^* =
	\left(C_{k_e^j}^{k_e^j+k_i^j}\right)^{-1}
	\sum_{i=1}^{C_{k_e^j}^{k_e^j+k_i^j}}
	\Delta_{inh}^j\Big(\mathcal{P}(\tau(\theta_K^{i,j}) + \tau_j)\Big),
	\label{eq:exp_inh_rt2}
\end{equation*}
where the $^*$ symbol next to the expected value denotes the fact that we used
the variable phase velocity approach. The definition of the phase deviation
$D$ in this equation reflects the case of heterogeneous synapses:
\begin{equation*}
	D = \sum_{j=1}^{N_{syn}} \left(
	\sum_{l=1}^{k_e^j} \Delta_{exc}^j(\theta(t_{l,j}^e)) +
	\sum_{m=1}^{k_i^j} \Delta_{inh}^j(\theta(t_{m,j}^i)) \right).
	    \label{eq:ph_dev2}
\end{equation*}
Obtaining the equivalent PRC proceeds as before. Let 
$b_j = \phi + \Delta_{exc}^j(\phi) + \omega d$; we now have:
\begin{equation*}
	E(\Delta_{inh}^j(b_j + D)|\phi) \approx
	\sum_{k_e^j=0}^\infty \sum_{k_i^j=0}^\infty
	\left[ \frac{(r_e^j d)^{k_e^j}}{k_e^j!} \text{e}^{-r_e^j d} \right]
	\left[ \frac{(r_i^j d)^{k_i^j}}{k_i^j!} \text{e}^{-r_i^j d} \right]
	E(\Delta_{inh}^j(b_j+D)|\phi,k_e^j,k_i^j,RT)^*,
	\label{exp_inh2}
\end{equation*}
where $r_e^j$ and $r_i^j$ are the excitatory and inhibitory firing rates 
of the $j$-th input. Since we are considering the case of
feedforward inhibition, we'll have $r_e^j = r_i^j$.
The fourth version of the equivalent PRC is:
\begin{equation}
	\Delta_{eq}^j(\phi) = \Delta_{exc}^j(\phi) +
        E(\Delta_{inh}^j(b_j + D)|\phi).
	\label{eq:eqprc_4}
\end{equation}
As would be expected, this PRC has the same limitations as the second
version of the equivalent PRC, and those limitations can be surmounted
using the same approach that led form the second to the third equivalent
PRC. The phase deviation when using the equivalent synapses is now:
\begin{equation*}
	D_{eq} = \sum_{j=1}^{N_{syn}} 
	\sum_{l=1}^{k_e^j} \Delta_{eq}^j(\theta_{eq}(t_{l,j}^e)).
	    \label{eq:eqph_dev2}
\end{equation*}
If at time $t$ an excitatory input is received by the $j$-th synapse,
the expected phase for the oscillator with feedforward inhibition and
heterogeneous synapses at time $t+d$ is: \\
$b_j + E(D|\phi,j) + E[\Delta_{inh}^j(b_j + D)|\phi,j]$. 
The notation $E(D|\phi,j)$ indicates the assumption that a shift of
magnitude $\Delta_{exc}^j(\phi)$ happens at time $t$.
In the case of the oscillator with the equivalent PRCs of equation
\ref{eq:eqprc_4}, the expected phase at time $t+d$ is: \\
$b_j + E(D_{eq}|\phi,j) + E[\Delta_{inh}^j(b_j + D)|\phi,j]$. 

Similarly to equation \ref{eq:xi1}, the difference in phase at time 
$t+d$ is:
\begin{equation*}
	\xi_j(\phi) = E(D|\phi,j) - E(D_{eq}|\phi,j).
\end{equation*}
 Let the functions $\Delta_{eq}^{j,(0)}$ come from equation
\ref{eq:eqprc_4}, for $j = 1,\dots,N_{syn}$.
A procedure to obtain the fifth version of the equivalent PRCs can be
as follows:
\begin{algorithmic}
	\FOR{$i$ = 1 to $M$}
	\FOR{$j$ = 1 to $N_{syn}$}
	\STATE $\xi^{(i)}_j(\phi) = E(D|\phi,j) - E(D_{eq}^{(i-1)}|\phi,j)$ 
	\STATE $\Delta_{eq}^{j,(i)}(\phi) = \Delta_{eq}^{j,(i-1)}(\phi) 
	       + \epsilon \ \xi^{j,(i)}_j(\phi) $
	\ENDFOR
	\ENDFOR
\end{algorithmic}
The value $E(D_{eq}^{(i-1)}|\phi,j)$ is the expected phase deviation 
calculated using the equivalent PRCs $\Delta_{eq}^{j,(i-1)}$. This 
calculation can be time consuming depending on the method used, so 
the algorithm can be modified by dividing the $N_{syn}$ equivalent
PRCs into $N_{batch}$ batches. Then, for each batch we use the same
$\xi^{j,(i)}_j$ in order to update the PRCs.

\begin{figure}
	\includegraphics[width=4in]{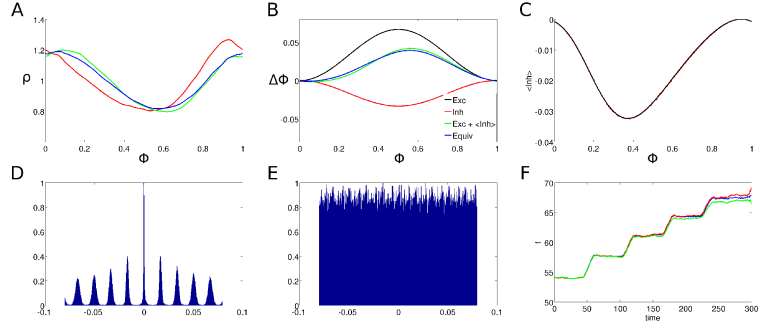}
\caption{
	{\bf Performance of the 2nd and 3rd Eq. PRC versions (homogeneous
	inputs) with 1 ms delay.}
These figures compare the results of simulating an oscillator with
feedforward inhibition against those of an oscillator with the 
equivalent PRC when the feedforward delay is 1 ms. The amplitudes
for the excitatory and inhibitory PRCs are 
$a_{exc} = 2/30$, and $a_{inh} = 1/30$ respectively. The input rate
is 600 Hz.
A: stationary phase PDFs for the oscillator with feedforward inhibition
(red), with the second version of the equivalent PRC (green), and
with the third version of the equivalent PRC (blue).
B: Phase resetting curves. Black=excitatory, red=inhibitory,
green=second Equiv. PRC, blue=third Equiv. PRC.
C: Expected value of the inhibition as a function of the phase when the
initial excitatory input arrives. The red curve comes from simulating
the oscillator with feedforward inhibition, and the black curve comes
from the approximation in equation \ref{eq:exp_inh_rt}.
D: Cross-correlogram between the output spike trains of the oscillators
with feedforward inhibition and with the third equivalent PRC. The
horizontal axis indicates time shift, and the vertical axis
the normalized spike count.
E: Cross-correlogram between the output spike train of the oscillator
with feedforward inhibition and a spike train with constant
interspike intervals and the same frequency.
F: Firing rate response of the oscillators with feedforward
inhibition (red), the second version of Equiv. PRC (green), and 
the third version of Equiv. PRC (blue) to five different levels of 
input rates. Input rates range from 240 Hz to 1200 Hz in constant 
increments of 240 Hz.
}
\label{result1_2_0}
\end{figure}

As before, I use computational simulations in order to validate the 
approximations obtained in this subsection. The first result shown
is that the third version of the equivalent PRC can already give accurate
results in replicating the response of an oscillator with feedforward
inhibition, especially when the feedforward delay is small. If this
delay is small enough, even the second version of the equivalent PRC
(equation \ref{eq:eqprc_2}) can give reasonable results. This is seen in figure 
\ref{result1_2_0}, where the feedforward delay is taken to be
1 ms. In panel A of figure \ref{result1_2_0}, it is seen that the
matching of stationary phase PDFs is not as good as in the previous
case, when the formulas where derived for this purpose. This is to
be expected; for each excitatory input the oscillator with feedforward
inhibition advances its phase, whereas the oscillator with the 
equivalent PRC experiences a shift in phase that includes an estimate 
of future inhibition. On the other hand, when tracking the timing of
output spikes, the oscillator with the third version of the PRC 
may outperform the oscillator with a phase PDF-matching PRC, as long
as the feedforward delay is small. This can be observed in panel
D of figure \ref{result1_2_0}, where the normalized cross-correlogram
shows a large similarity between the output spike trains. It could be
argued that this similarity in the cross-correlograms arises from both
spike trains having the same frequency. To counter this argument, I 
also show in panel E the cross-correlogram between the spike train with 
delayed inhibitory inputs, and a spike train with constant
interspike intervals and the same frequency.

Another relevant result is in panel C of figure
\ref{result1_2_0}. This panel shows a close agreement between the 
expected inhibitory shift at time $t+d$ obtained from simulations 
(red curve), and from equation \ref{eq:exp_inh_rt} (black curve),
which approximates the expected inhibition by assuming that
during the inhibitory delay all inputs will arrive at regular
intervals.
	
Reported values of delay between an excitatory input and the
corresponding feedforward inhibition are usually in the 1-2 ms range
\cite{mittmann_feed-forward_2005,mittmann_linking_2007},
and have even been described as non-existent
\cite{blot_ultra-rapid_2014}.
It is nevertheless relevant to test whether the 
formulas of this paper can be still applicable when the
feedforward delay is not as short. The approaches taken here to derive 
the equivalent PRCs should show their shortcomings as the delay
and the input firing rate are increased. With this in mind the
simulations in this paper --with the exception of the one in
figure \ref{result1_2_0}-- use a delay of 5 ms, larger than
reported values, but still biologically plausible.

\begin{figure}
	\includegraphics[width=4in]{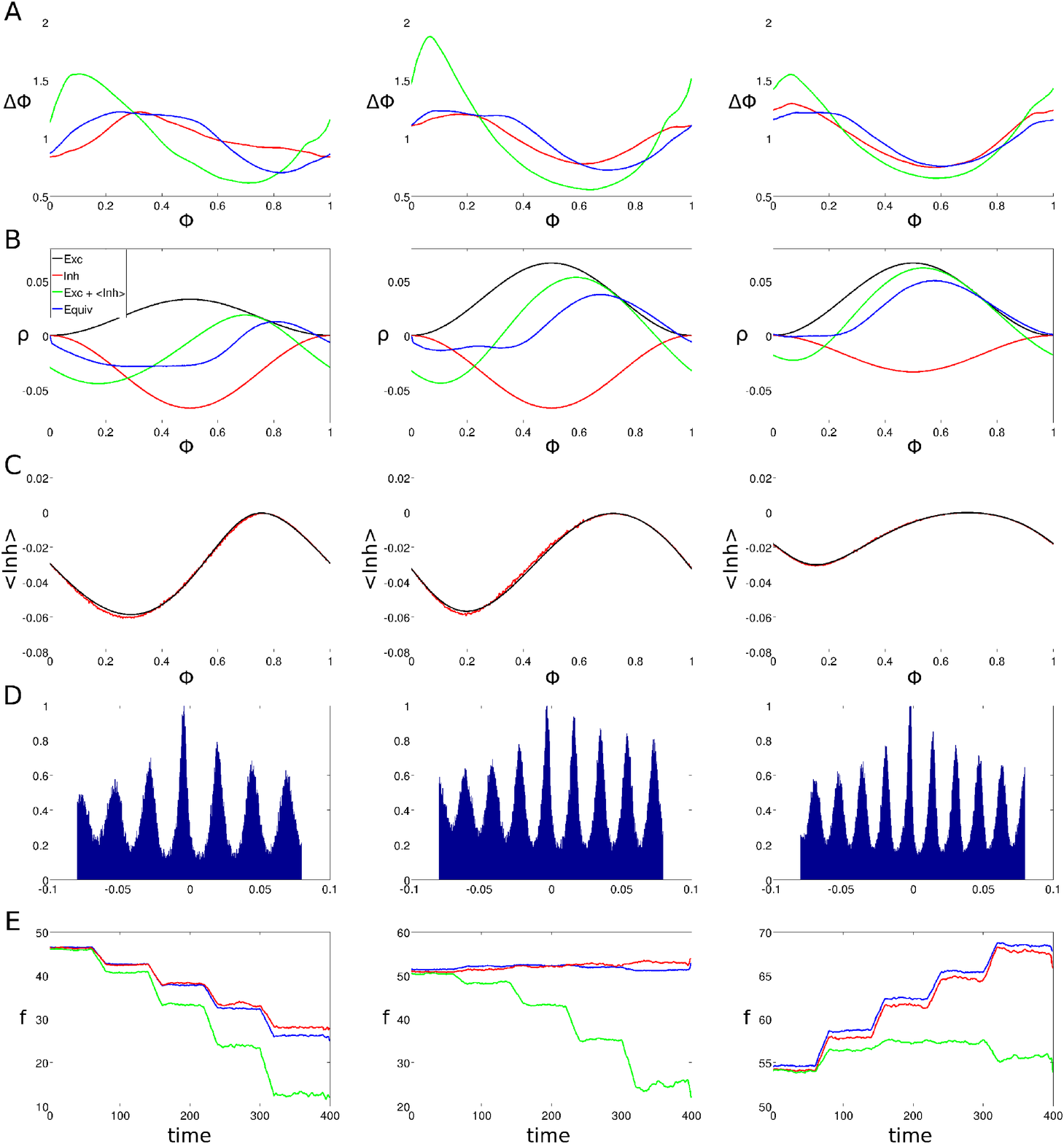}
\caption{
{\bf Performance of the 2nd and 3rd Eq. PRC versions (homogeneous inputs) with 
5 ms delay. }
Simulations for 3 different 
amplitude combinations of $\Delta_{exc}$ and $\Delta_{inh}$. Figures on the
left correspond to $a_{exc} = 1/30, \ a_{inh} = 2/30$. Figures on the
middle correspond to $a_{exc} = 2/30, \ a_{inh} = 2/30$, and Figures on the
right correspond to $a_{exc} = 2/30, \ a_{inh} = 1/30$. The input rate for
all simulations in A-D was $r = 600$ Hz. Other than the input amplitudes, the
only difference with the simulation of figure \ref{result1_2_0} is the
delay value of 5 ms.
A: stationary phase PDFs for the oscillator with feedforward inhibition
(red), with the second version of the equivalent PRC (green), and
with the third version of the equivalent PRC (blue).
B: PRCs used in the simulation. $\Delta_{exc}$
(black), $\Delta_{inh}$ (red), second version of $\Delta_{eq}$ (green),
third version of $\Delta_{eq}$ (blue). 
C: Expected value of the inhibition as a function of the phase when the
initial excitatory input arrives. The red curve comes from simulating
the oscillator with feedforward inhibition, and the black curve comes
from the approximation in equation \ref{eq:exp_inh_rt}.
D: Cross-correlogram between the output spike trains of the oscillators
with feedforward inhibition and with the third equivalent PRC. The
horizontal axis indicates time shift, and the vertical axis
the normalized spike count.
E: Firing rate response of the oscillators with feedforward
inhibition (red), the second version of Equiv. PRC (green), and 
the third version of Equiv. PRC (blue) to five different levels of 
input rates. Input rates range from 240 Hz to 1200 Hz in constant 
increments of 240 Hz.
}
\label{result1_2}
\end{figure}

Figure \ref{result1_2} illustrates simulations done with the second
and third versions of $\Delta_{eq}$ and a feedforward delay of
5 ms. It is evident in panel E that the second version of the
PRC becomes incapable of matching the firing rate of the oscillator
with feedforward inhibition as the input rates become larger. On
the other hand, the oscillator with the third version of 
$\Delta_{eq}$ still displays similar behaviour to the oscillator
with feedforward inhibition.

\begin{figure}
	\includegraphics[width=4in]{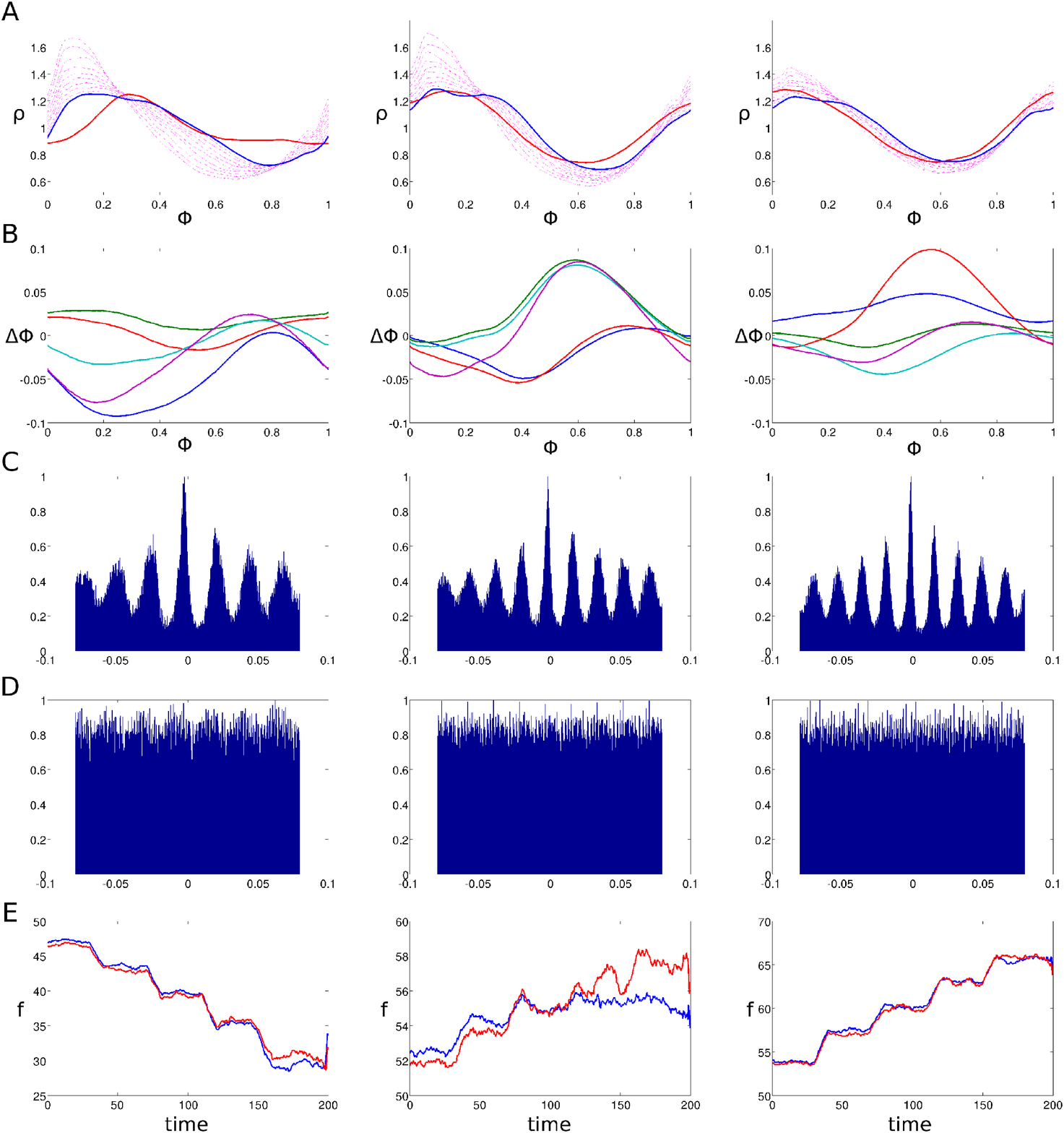}
\caption{
{\bf Performance of the fifth Eq. PRC version (heterogeneous inputs) 
with 5 ms delay.}
Simulations for 3 different combinations of mean PRC amplitudes. 
For each simulation a total of $N_{syn}=60$ different inputs 
were used, each with randomly chosen amplitudes $a_{exc}^j$ and $a_{inh}^j$
for $\Delta_{exc}^j$ and $\Delta_{inh}^j$. The firing rate for all
inputs was $r_j = 10$ Hz.
Figures on the left correspond to a simulation where the mean of the PRC
amplitudes is $<a_{exc}^j> = 1/30$, $<a_{inh}^j> = 2/30$. Figures on the
middle correspond to $<a_{exc}^j> = 2/30$, $<a_{inh}^j> = 2/30$, and figures 
on the right correspond to $<a_{exc}^j> = 2/30$, $<a_{inh}^j> = 1/30$. 
A: stationary phase PDFs for the oscillator with feedforward inhibition
(red), and with the fifth version of the equivalent PRCs (blue). 
Magenta dash-dotted lines indicate the phase PDF corresponding to intermediate
iterations in the iterative procedure used.
B: PRCs used in the simulation. Different colored lines show the equivalent
PRCs for the first 5 synapses.
C: Cross-correlogram between the output spike trains of the oscillators
with feedforward inhibition and with the equivalent PRCs. The
horizontal axis indicates time shift, and the vertical axis
the normalized spike count.
D: Cross-correlograms between the output spike train of the oscillator with
feedforward inhibition and a regular spiking oscillator matching its mean
frequency.
E: Firing rate response of the oscillators with feedforward
inhibition (red), and  the equivalent PRCs (blue) to five different levels of 
input rates. Input rates range from 240 Hz to 1200 Hz in constant 
increments of 240 Hz.
}
\label{result1_3}
\end{figure}

The fifth version of the equivalent PRC was also tested, using 60
different inputs, each one with different amplitudes for its excitatory
and inhibitory PRCs. The result of the simulations is illustrated
in figure \ref{result1_3}. As can be seen from this figure, and from 
figure \ref{result1_2}, the performance for oscillators with homogeneous 
and heterogeneous inputs is very similar.

\section{Computational explorations of synchronization with delayed inhibition}

The consequences of stochastic synchronization of Purkinje cells have not been
explored before. As an initial approach to this endeavor, I explored how
various factors could affect the synchrony of uncoupled oscillators receiving
correlated inputs and a delayed inhibitory stimulus for each excitatory stimulus.
The insight that there exists an equivalent PRC governing the response of the
oscillators means that previous studies on stochastic synchronization (where
each input is described with a single PRC) should be applicable in this case. 
As such, this exploration is guided by those studies. Stochastic synchrony is a
complex and incompletely understood subject, but it is the author's opinion that
a large part of the existing results can be intuitively understood in terms of
3 factors: input correlation, heterogeneity, and PRC shape. Each of these 3 factors
will be manipulated in this section. In the cases of input correlation and heterogeneity
it will be shown that these factors can be manipulated without altering the PRCs (which in this
case correspond to synaptic and excitability effects), but by altering instead the 
input sources that have a high firing rate. In this way it will be shown that different
input patterns in the parallel fibers can affect the degree of synchronization
in the Purkinje cells, without necessarily affecting their firing rate.

For each factor controlling stochastic synchrony, I will also mention some
possible ways by which plasticity could affect it. Climbing fiber
activity has been shown to mediate plasticity in several types of
cerebellar cortical synapses \cite{hansel_beyond_2001}, and in this paper I 
focus on plasticity mediated by CF activity in 3 synapse types: 
parallel fiber to Purkinje cell (PF-PC), molecular
layer interneurons to Purkinje cells (MLI-PC), and parallel fiber to
molecular layer interneurons (PF-MLI).

\subsection{Details of the simulations}

All simulations used 20 oscillators with frequencies
drawn from a Gaussian distribution with mean of 50 Hz, and a standard 
deviation of 1 Hz. Each oscillator received 60 separate inputs, selected from
a pool of different Poisson point processes. The effect of 
feedforward inhibition was simulated by applying an inhibitory input 5 ms
after each excitatory input; this delay was chosen to illustrate that the
effects described in this paper are present despite significant delays in the
inhibition. As previously discussed, the effect of each excitatory and inhibitory
input was described by separate PRCs, with excitatory PRCs being positive
or null for all phases, and inhibitory PRCs being negative or null.
The shape of all PRCs was the sinusoidal (positive or negative) used
in the previous section.
All simulations used an exact integration method, where the phases of
all oscillators were updated whenever an input was received or one of them
spiked.

Three different measures were used in order to test the synchrony of the
20 oscillators.
The first and most significant measure was obtained by simulating
a DCN neuron receiving all the spikes produced by the 20 oscillators. 
Neurons in the DCN
tonically produce action potentials, and have powerful depolarizing currents
that are triggered by hyperpolarization \cite{jaeger_mini-review:_2011}. These
features were included in a simple model of a DCN neuron consinsting of an
oscillator whose frequency was raised by deep enough hyperpolarizations but
relaxed back to baseline levels for moderate amounts of inhibition. 
The phase of the DCN neuron, represented as $\theta_{DCN}$ had an angular
frequency $\omega_{DCN}$, and obeyed the equation
\begin{equation*}
	\dot \theta_{DCN} = \omega_{DCN} - \sum_i a \delta(t-t^i) \ .
\end{equation*}
The value $t^i$ represents the arrival time of the $i$-th input,
$\delta()$ is the Dirac delta function, and $a$ is a positive value
representing the amplitude of the inhibitory inputs.
The value $\theta_{DCN}$ is not an ordinary phase, as it is allowed 
to remain in the interval $[-2,1]$. Whenever the phase reaches the value 1 
it is reset to the value 0, but the negative inputs can create negative values
that saturate at -2.
Whenever $\theta_{DCN}$ goes below the threshold value $\theta_{thr} = -0.5$
the frequency $\omega_{DCN}$ increases according to
\begin{equation*}
	\dot \omega_{DCN} = \frac{ \omega_{max}-\omega_{DCN}}
	{1 + e^{\xi(t_{thr}-t+t_{off})}},
\end{equation*}
where $\omega_{max}$ is the maximum angular frequency, $t_{thr}$ is the time
when the phase last crossed the threshold, and $\xi, t_{off}$ are positive 
constants. On the other hand, when the frequency is above the threshold
$\theta_{thr}$, $\omega_{DCN}$ obeys
\begin{equation*}
	\dot \omega_{DCN} = \frac{ \omega_{base}-\omega_{DCN}}
	{1 + e^{\xi(t_{thr}-t+t_{off})}},
\end{equation*}
where $\omega_{base}$ is the angular frequency that would take place in the
absence of inputs. Notice that in the absence of threshold crossings these
two equations asymptotically become $\dot \omega_{DCN} = \omega_{base/max}
- \omega_{DCN}$.
Volleys of synchronized spikes from the 20 oscillators would create deep
hyperpolarizations (negative phases) in the DCN cell, and afterwards provide a 
time period with low inhibition, during which the DCN cell could spike. The result 
was that that for fixed firing rates in the input, the output firing rate of the 
DCN cell increased as a function of the input synchrony.

The second synchrony measure used was the circular variance 
\cite{allen_automated_1991}. In order
to obtain this measure, each time one of the oscillators spiked, the
phases of the other 19 oscillators were obtained. This resulted in a 
large sample of phases $\phi_i$, $i = 1,\dots,N$ which should be clustered 
near the values 0 or 1 in the case when the oscillators spike close to one 
another. Since the phases are periodic, in order to average them they are 
represented as complex numbers with unit norm. If the number corresponding
to $\phi_i$ is $s_i$, then $\mathbb{R}\{s_i\} = \cos(\phi_i)$,
$\mathbb{I}\{s_i\} = \sin(\phi_i)$, and the average of these numbers is
$\bar s = \frac{1}{N} \sum_{i=1}^N \left( \cos(\phi_i) + i \sin(\phi_i)
\right)$. The circular variance is defined as $1 - \| \bar s \|$.

The third synchrony measure comes from \cite{schreiber_new_2003}, 
and approximates
the average correlation among spike trains. In short, spike trains are 
convolved with a Gaussian filter to obtain their continuous version, and
the correlation among two spike trains comes from the normalized inner 
product of their continuous versions. The synchrony measure comes from the
mean of these correlations for all pairs of spike trains. In the author's
implementation of this measure, due to the widths of the Gaussian filters
used, the synchrony measure obtained from a group of spike trains
$A$ can be compared to the synchrony of a group of spike trains $B$ only
when the trains in $A$ and $B$ have similar mean firing rates.

The values for synchrony measures and average firing rates in the bar graphs
of all figures come from averaging the results of 100 simulations.
The simulations and all other calculations were performed in Matlab
(www.mathworks.com). Source code is available upon request.

\subsection{The effect of input correlation}
Previous studies show that the more correlated the inputs are, the greater 
the potential for stochastic synchronization of their targets. In other
words, for any two oscillators larger input correlations cause a peak 
in the distribution of their phase difference 
\cite{nakao_synchrony_2005,marella_class-ii_2008,zhou_impact_2013}. 

\begin{figure}
	\includegraphics[width=6in]{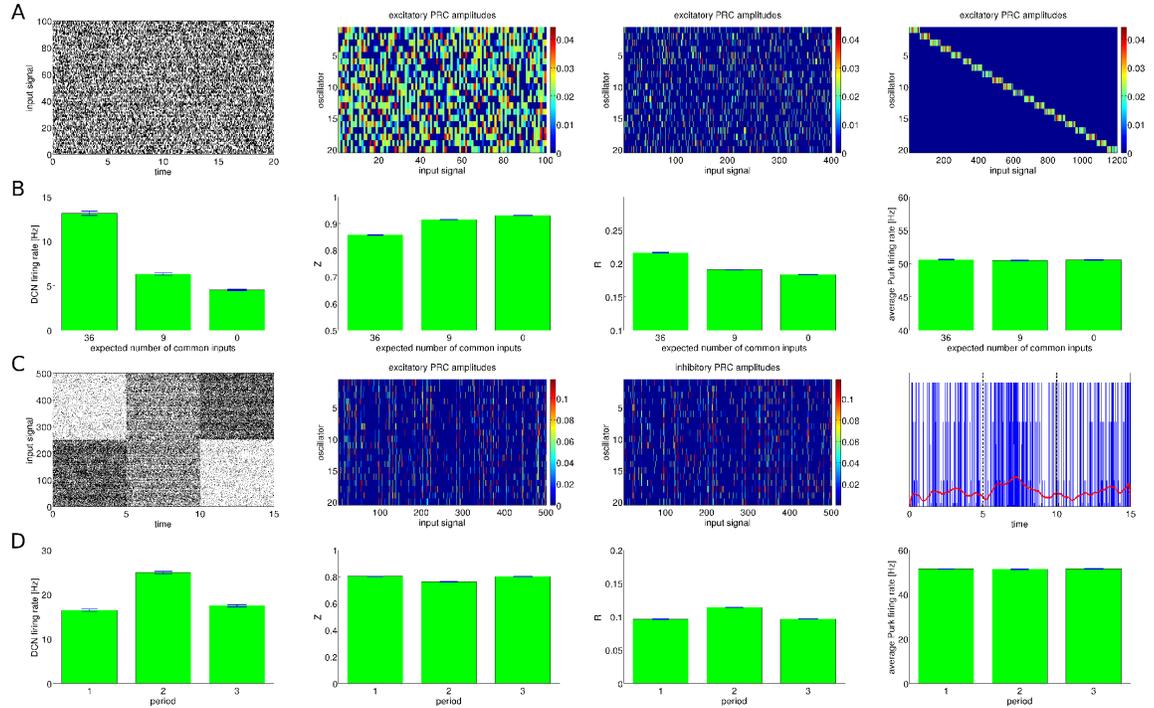}
\caption{
{\bf Effect of input correlation on synchrony of multiple oscillators}  
A: (Far left) Spike raster for 100 input signals during 20 seconds of simulated
time. (Middle left) excitatory connection weights from the 100 input signals to 
the 20 oscillators. These weights correspond to the amplitudes of the 
excitatory PRCs; the amplitudes of inhibitory PRCs are not shown. 
(Middle right) excitatory connection weights from 400 input signals to the 
20 oscillators. (Far right) excitatory connection weights from 1200 input
signals to the 20 oscillators in the case of uncorrelated inputs.
B: DCN firing rate, circular variance, mean correlation, and Purkinje oscillator
average firing rate for simulations with 3 different values for the expected
number of common inputs, corresponding to connection matrices like the
ones in panel A.
C: Representative simulation displaying the effect of more high-frequency inputs
being shared by the oscillators. (Far left) spike raster for the 500 inputs
during the 3 periods of the simulation. (Middle left) excitatory connection 
weights. (Middle right) inhibitory connection weights. (Right) output spike
train of the DCN oscillator convolved with a thin Gaussian. The red trace is
a smoothing of this signal, which allows to see the fluctuations in the
firing rate.
D: DCN mean firing rate, circular variance, mean correlation, and Purkinje 
oscillator firing rate for the 3 periods shown in C, averaged over 100 simulations.
}
\label{result2_2}
\end{figure}

The simulation results in panels A and B of figure \ref{result2_2} illustrate
this. There were 3 simulated trials, each one with a different level of
input correlation among the 20 oscillators. In order to change the level
of input correlation on each trial, the number of possible input sources
(each a Poisson point process) was increased, so that with each oscillator
receiving input from 60 randomly chosen sources, the expected number of
common inputs decreased. The first figure in panel A of figure \ref{result2_2}
shows the spike raster of the 100 input sources from the first trial. The
other 3 figures show the amplitudes of the excitatory PRCs depending on the
oscillator and the input source. An amplitude of zero (denoted by the blue
color) means that the oscillator is not connected to the input source, so
for each row there are 60 columns with a color other than dark blue, corresponding
to values chosen randomly from a uniform distribution. This makes
explicit that the expected number of common inputs among different oscillators
changes between trials, from 36 to 9 to 0.
In panel B it is shown that increasing input correlation increases the
synchrony of the oscillators as measured by the circular variance and mean
correlation described previously. The firing rate of the simulated DCN
cell also increases when there are more common inputs. The figure on the
far right of panel B shows that changing the number of common inputs under
these conditions doesn't have an effect on the firing rate of the oscillators.

The next simulation involving input correlation involves a single trial
where the input sources have 3 different firing patterns, and is shown on
panels C and D of figure \ref{result2_2}. It is relevant to ask whether 
activating a particular combination of cells in the granule-cell layer could 
increase the response in the DCN cells of its microcomplex, not because of 
the average weight of their synapses onto Purkinje cells, but because they
are common to many of those Purkinje cells. Mossy fiber LTP
\cite{hansel_beyond_2001} promotes a population code, in which sensory 
stimuli can be represented by particular combinations of granule cells 
with a high firing rate, and Golgi feedback inhibition ensures that these 
combinations do not remain static \cite{dangelo_timing_2009}, so a candidate
code for the information in the granule cell layer could be combinations of
cells spiking faster.

The simulations shown in figure \ref{result2_2}, panels C and D have 
500 inputs, and at all times half of them have a low
firing rate (5 Hz), and half of them have a high firing rate (40 Hz). 
The PRC amplitudes for all inputs were randomly chosen from a uniform 
distribution in the interval [0, 0.06].
Numbering the 500 inputs, 250 are even and 250 are odd. 30 odd inputs
were chosen to be received by every oscillator. For each oscillator,
its remaining 30 inputs were randomly chosen from the other 470 inputs.
The simulation then proceeded in 3 stages, each lasting 5 seconds. In the
first stage, inputs 1 through 250 had a high firing rate, and the last
250 inputs had a low firing rate. In the second stage the 250 odd inputs had
a large firing rate, and the even inputs had a low firing rate. In the
third stage the last 250 inputs had a high firing rate, and the first
250 inputs had a low firing rate. The result is that the synchrony of
the oscillators increased during the second stage, when the 30 common
odd inputs had a high firing rate. The average firing rate of all 
oscillators was the same during the 3 stages.

We can ask what role plasticity could play in shaping the input correlation.
The majority of PF-PC synapses seem to be silent 
\cite{isope_properties_2002,wang_quantification_2000}. Silent synapses can
become active, and active synpses can become silent 
\cite{ekerot_parallel_2003}, which suggests
that Purkinje cells are continuously selecting their inputs. 
When there is no climbing fiber activity, sensory-evoked high-frequency 
input creates LTP in the PF-PC synapse 
\cite{lev-ram_new_2002,coesmans_bidirectional_2004,wang_long-term_2009,
schonewille_purkinje_2010}. The cell seems to select those 
inputs that are strong and don't happen near complex spikes. When there
are complex spikes, conjunctive CF and PF activity leads to LTD in the
PF-PC synapse \cite{ito_climbing_1982,ito_synaptic_1993,
mittmann_linking_2007,hirano_long-term_2013}.
This is balanced with LTD in the MLI-PC synapse \cite{mittmann_linking_2007}.
The cell tends to ignore inputs that happen at the time of complex spikes.
This suggests that input correlation may be weakened for inputs that
coincide with complex spikes. The effect of this on the DCN firing rate
would be similar to the effect attributed to conjunctive LTD in the 
PF-PC synapse, suggesting that these two mechanisms could be complementary.

\subsection{The effect of heterogeneity}
Physiological heterogeneity can disrupt stochastic synchronization
\cite{smeal_phase-response_2010,markowitz_rate-specific_2008,
burton_intrinsic_2012,zhou_impact_2013}. This heterogeneity can
present itself in the firing rates of the oscillators, and in the
variety of responses to different inputs. In here I focus on the
heterogeneity of responses, characterized as PRC heterogeneity.

If we have uncoupled oscillators, and the response to inputs for each one is
characterized with a separate PRC, then heterogeneity in those PRCs
can limit the amount of stochastic synchronization 
\cite{burton_intrinsic_2012,zhou_impact_2013}. Small-amplitude type II
PRCs are more susceptible to this effect of heterogeneity than 
large-amplitude PRCs \cite{burton_intrinsic_2012}. In the case of low
input correlations, two heterogeneous oscillators may synchronize better
than two homogeneous ones, but only in the case when the two homogeneous
are ``bad synchronizers,'' and the heterogeneous ones include a ``good
synchronizer'' \cite{zhou_impact_2013}. In general, the best case for
stochastic synchrony is when the PRCs are similar, and their shape is
conducive to synchronization, as in the case of type II PRCs (discussed
in the next subsection).

\begin{figure}
	\includegraphics[width=6in]{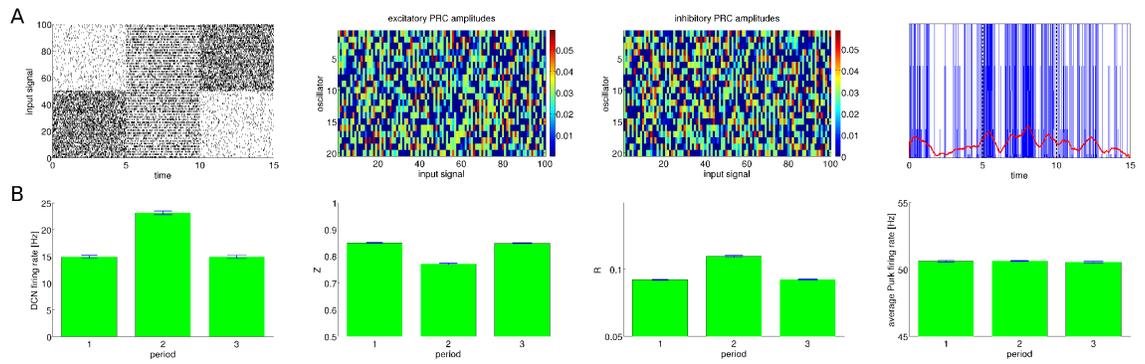}
\caption{
{\bf Effect of PRC homogeneity on synchrony of multiple oscillators.}  
A: Representative simulation displaying the effect of homogeneous synapses
being targetted by the high-frequency inputs.
(Far left) spike raster for the 100 inputs
during the 3 periods of the simulation. (Middle left) excitatory connection 
weights (PRC amplitudes). (Middle right) inhibitory connection weights. Although
it is hard to appreciate in this figure, for each odd input its inhibitory and
excitatory weights are the same.
(Right) output spike train of the DCN oscillator convolved with a thin 
Gaussian. The red trace is a smoothing of this signal, which allows to see 
the fluctuations in the firing rate.
B: DCN firing rate, circular variance, mean correlation, and average Purkinje 
oscillator firing rate for the 3 periods shown in C, averaged over 100 
simulations.
}
\label{result2_1}
\end{figure}

Figure \ref{result2_1} shows the result of a simulation in which increasing 
PRC homogeneity can increase oscillator synchrony and the DCN firing rate, 
and this is caused by shifting to a particular firing pattern of activity in 
the inputs. The firing pattern that increases the DCN activity does so not 
because it targets weaker or more inhibitory synapses, but because it
targets homogeneous PRCs.

Panels A and B of figure \ref{result2_1} show a single representative simulation
showing the effects of targeting homogeneous synapses, in a manner similar to
panels C and D of figure \ref{result2_2}. For this simulation,
there is a pool of 100 possible input sources. All the odd inputs have the
same excitatory and inhibitory PRC amplitudes. The even inputs have PRC
amplitudes randomly chosen from a homogeneous distribution whose mean is
the amplitude of odd PRCs. For the odd inputs, the excitatory and inhibitory
amplitudes were chosen to be equal, because in this case the equivalent
PRC has a proper shape for synchronization given the shapes and 
delay used (figure \ref{result1_2} B). Each oscillator randomly receives
30 odd and 30 even inputs. 
The simulation proceeds in 3 periods, each lasting 5 seconds, as was done
previously. In the first period the first 50 inputs have high firing rates
(40 Hz), and the last 50 inputs have low firing rates (5 Hz). During the
second period the odd inputs have high firing rates, and the even inputs
have low firing rates. In the last period the last 50 inputs have high
firing rates, and the first 50 inputs have low firing rates. 
The result is that during the second period the DCN oscillator reliably 
increases its firing rate. The average firing rate of all Purkinje oscillators
remains the same during the 3 periods, so the increase in DCN firing rate
is due to an increase in synchrony. Panel B of figure \ref{result2_1} shows
this effect for the average measures from 100 simulations.

We can also discuss the effects of plasticity on input heterogeneity, although
unlike input correlation, in this case it is case it is hard to reach
a hypothesis. It was mentioned in the previous subsection 
that high-frequency activity elicited by sensory stimulation can lead to
LTP in the PF-PC synapse. Moreover, climbing fiber activity that
is not paired with PF activity leads to LTP in the MLI-PC synapse, a
phenomenon known as rebound potentiation \cite{kano_synaptic_1992}. 
The PRC amplitudes depend on
the synapses, and on intrinsic excitability of the cell. Potentiation 
of the synapse could potentially lead to saturation in the PRC
amplitudes, so that all synapses with saturated amplitudes have
similar PRCs, thus decreasing heterogeneity. In the case when there
is conjunctive CF-PF activity both the PF-PC and MLI-PC experience
LTD. If the LTD in these two synapses is balanced, the heterogeneity
could be maintained. It is thus hard to speculate whether climbing
fiber activity alters the amount of heterogeneity, particularly when
there are also other forms of plasticity whose effect on the shape
of the effective PRC is not entirely clear.

\subsection{The effect of PRC shape and amplitude}

In general, most PRC shapes can produce some degree of stochastic
synchrony for small correlated inputs 
\cite{nakao_synchrony_2005,nakao_noise-induced_2007}. 
The particular shape of the PRC,
however, can have a considerable effect on the degree of synchronization.
Oscillators with type II PRCs, which are positive for late
phases and negative for early phases \cite{hansel_synchrony_1995}, 
tend to synchronize better when receiving correlated inputs 
\cite{marella_class-ii_2008,galan_stochastic_2007,burton_intrinsic_2012}. 
For small amplitude inputs the optimal shape for synchronization is a 
negative sinusoidal wave \cite{abouzeid_type-ii_2009},
but this shape may vary for different input amplitudes, and some shapes
may be desynchronizing \cite{hata_optimal_2011}.

In the case of feedforward inhibition, the balance between excitatory and
and inhibitory PRC amplitudes affects the shape of the equivalent PRC. 
If the excitatory and inhibitory PRCs have 
the same amplitude and a single peak, their equivalent
PRC tends to be of type II (depending on the peak locations and the duration
of the feedforward delay), and increasing this amplitude will in general
be favorable for synchronization. When the excitatory and inhibitory
PRCs have different amplitudes, increasing or decreasing those amplitudes
in the same proportion will have an effect in the firing rate of the
oscillator, so the effect of Purkinje PRC amplitude on the DCN firing rate 
will be largely due to changes in the firing rate of Purkinje cells. 

\begin{figure}
	\includegraphics[width=6in]{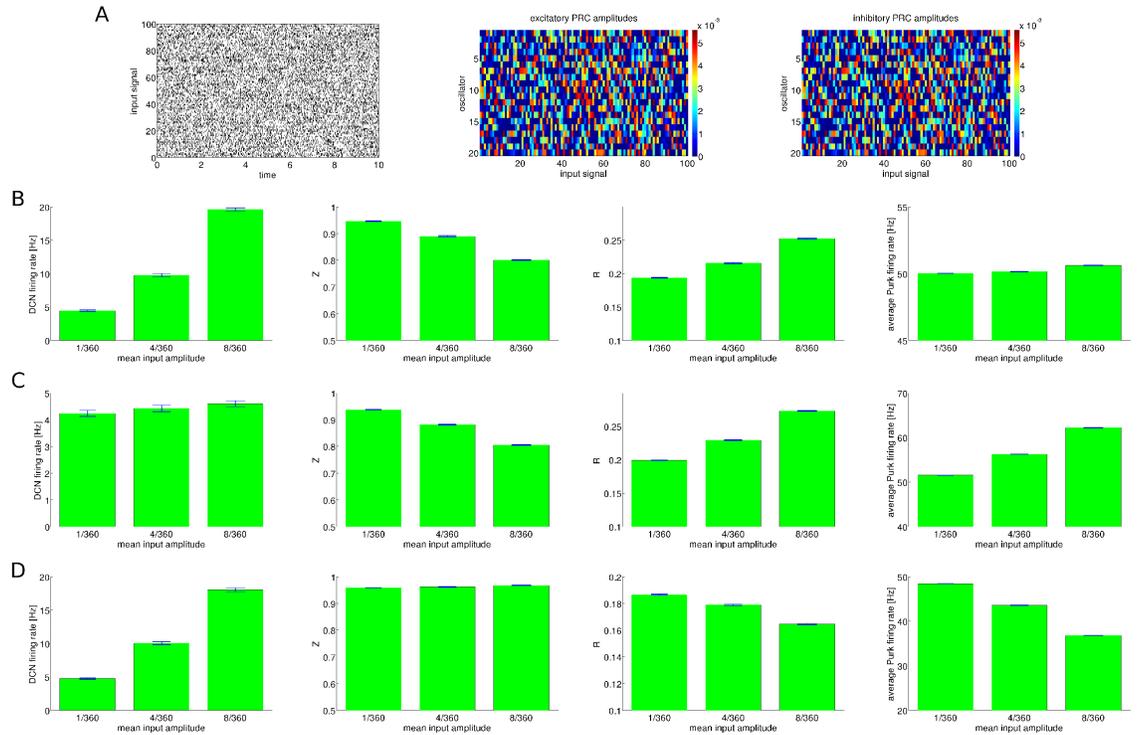}
\caption{
{\bf Effect of PRC shape and amplitude.}  
A: (Left) spike raster showing 100 inputs with constant firing rate; 
similar inputs were used for all simulations in this figure.
(Center and right) PRC amplitudes for the case of balanced excitation
and inhibition. Notice that amplitudes are the same for both cases.
B: DCN firing rate (far left), circular variance (middle left), 
mean correlation (middle right), and average Purkinje oscillator 
firing rate (far right) for 3 different values of mean PRC amplitude
in the case of balanced excitation and inhibition.
C: DCN firing rate (far left), circular variance (middle left), 
mean correlation (middle right), and average Purkinje oscillator 
firing rate (far right) for 3 different values of the mean PRC 
amplitude in the case when excitatory PRC amplitudes are twice as 
large as inhibitory PRC amplitudes.
D: DCN firing rate (far left), circular variance (middle left), 
mean correlation (middle right), and average Purkinje oscillator 
firing rate (far right) for 3 different values of the mean PRC 
amplitude in the case when inhibitory PRC amplitudes are twice as 
large as excitatory PRC amplitudes. The seemingly significant
decrease in mean correlation as opposed to the very slight 
change in circular variance is an artifact of comparing the
synchrony of spike trains with very different firing rates
(see Materials and Methods).
In this case, circular variance is the appropriate measure.
}
\label{result2_3}
\end{figure}

The simulations of figure \ref{result2_3} are intended to illustrate how
the effect of changing the PRC amplitudes depends on the shape of the
equivalent PRC, which is determined by the amplitudes and shapes of the
excitatory and inhibitory PRCs. The shapes of the excitatory and inhibitory 
PRCs are not manipulated, but the ratio of their amplitudes is set to be either
1:1, 2:1, or 1:2, which produces 3 different shapes for the equivalent PRC. 
For each of these 3 ratios, I run simulations for 3 different values of PRC
amplitudes. Each ratio corresponds to one of the B, C, D panels, and each
value of amplitude corresponds to one of the bars in the individual plots.

Panel A in the first row of figure \ref{result2_3} shows input rasters and
PRC amplitude matrices for a simulation with 
constant firing rates for 100 input sources. The randomly chosen amplitude 
matrices shown in panel A correspond to the case of balanced excitation 
and inhibition, and therefore these matrices are identical. In the case for
the other two ratios of excitation to inhibition, the matrices are scaled
versions of one another.
Panel B in the second row shows the effect of changing
the PRC amplitudes in the case of balanced excitation and inhibition.
As can be observed, a proportional change of PRC amplitudes has a very
small effect on the firing rate of the Purkinje oscillators, but the 
synchrony and DCN firing rate increase with the amplitude of the
PRCs, as would be expected given that the shape of the equivalent PRC
in this case tends to be of type II. This outcome changes when the amplitudes 
of excitatory PRCs are twice as large as those of the inhibitory PRCs. This case 
can be seen in Panel C of figure \ref{result2_3}, where increasing
the PRC amplitudes proportionally significantly increases both synchrony
and Purkinje firing rate, and with these two factors opposing each 
other the DCN firing rate only shows a very small increase. It should
be remembered that the author's implementation of the mean correlation 
measure R is not a good way to compare the synchrony of different trials
when they have different mean firing rates (the case in panels C and D),
so the Z measure should be used instead, which shows smaller values for
larger levels of synchrony.
A different outcome happens when the inhibitory PRCs are twice as large
as the excitatory PRCs and the amplitudes are increased while 
maintaining these proportions. Panel D shows that in this scenario
the synchrony barely decreases as the amplitudes grow, but the
Purkinje firing rates decrease significantly, resulting in an increase
of the DCN firing rate. This last scenario where the synchrony's 
effect is small replicates the hypothesis about the role of
conjunctive LTD in the PF-PC synapse in the Marr-Albus-Ito model.
To the author's knowledge, no cerebellar model has explicitly considered the
possibility of synchrony effects such as those in panels B and C.

Clearly, plasticity in the synapses of Purkinje cells will have an
effect on the shape and amplitude of their equivalent PRCs. 
The aforementioned LTP in the PF-PC synapse created by PF activity
in the absence of CF activity will tend to increase the amplitude
of excitatory PRCs, and rebound potentiation will create
the corresponding effect for inhibitory PRCs. Conjunctive PF and
CF activity will reduce both excitatory and inhibitory PRC amplitudes. 
The effect of this change in amplitudes will depend on a various 
details, such as the balance in excitation and inhibition, the 
shapes of the PRCs, and the importance of Purkinje cell synchrony 
on DCN output relative to the importance of IPSP amplitude.
As long as these factors are not considered, intuitive predictions
about the role of plasticity in shaping DCN output could be
erroneous.

\section{Discussion}
I have shown that stochastic synchronization in the presence of feedforward
inhibition with relatively small delays can be studied with the standard
PRC methods using the equivalent PRC concept. 
Furthermore, I have suggested how stochastic synchronization 
in Purkinje cells of the cerebellum could be explored in terms of three
properties: input correlation, heterogeneity (of PRCs and firing rates), 
and PRC shape. 

\subsection{Two principles for studying stochastic synchronization 
with delayed feedforward inhibition}

There is  increasing evidence for the role of synchrony in the the code
of Purkinje cells' output \cite{person_synchrony_2012,de_zeeuw_spatiotemporal_2011}. 
For a long time it 
has been hypothesized that this synchrony arises due to mossy fiber input
\cite{bell_discharge_1969}, and in the case of simple spike synchronization
among Purkinje cells separated by several hundred micrometers along the
direction of parallel fibers this hypothesis may be correct
\cite{ebner_temporal_1981,heck_-beam_2007,wise_mechanisms_2010,
bosman_encoding_2010}. Modulation of firing
rates by common inputs can increase the amount of synchrony observed
in a cross-correlogram, but when synchrony increases while firing rates
are not being modulated, or when they are modulated in different directions,
then a more complete approach to studying synchrony is required. The
PRC of a cell is a compact representation of all the factors that affect
its input-driven synchronization, so the application of theoretical
advances in stochastic synchronization is a sensible approach to understand
the effect of inputs to Purkinje cells.  Studying this,
however, may not be straightforward given all the connections involved
in the cerebellar cortical loop. This paper suggests two
principles to begin this study. A first principle is that feedforward
inhibition can be conceptually removed by considering equivalent
PRCs. A second principle is that synchrony is propitiated by:
(i) similar firing rates in Purkinje cells, (ii) homogeneous
PRCs, (iii) correlated inputs, and (iv) type II equivalent PRCs.

In theory these two principles can be experimentally tested, but
some aspects appear technically challenging, such as manipulating the
input correlation of Purkinje cells, or the homogeneity of their
synapses. Detailed modeling is a viable alternative, although it is
hampered by unknown physiological data, such as the shape and frequency
dependence of inhibitory PRCs, statistical properties of some connections,
or a detailed understanding of how DCN cells integrate their input. 
Nevertheless, there are realistic community models \cite{bower_emergence_2013}
that have approached similar challenges 
(e.g. \cite{jaeger_no_2003,santamaria_feedforward_2007,
sudhakar_biophysical_2013,jaeger_mini-review:_2011}),
so their use along with experimental approaches is a promising future 
direction.

\subsection{Is the PRC model adequate?}

Purkinje cells are particularly complex, and their computational models tend to 
be mathematically intractable. The oscillator representation is a mathematically 
tractable model that allows to construct complicated hypotheses that may then
be addressed by physiological experiments and by detailed models. This extra
step in the modeling process is beneficial in the study of synchronization
because the effects of changing physiological parameters are not straightforward
in this case. Still, modeling Purkninje cells as oscillators and characterizing 
their inputs through a PRC entails a drastic reduction in complexity, which 
could obscure important details. 
Purkinje cells tonically fire simple spikes with frequencies ranging 
from ~30 to 150 Hz; these frequencies are modulated by afferent
and efferent inputs 
(e.g. \cite{holtzman_different_2006,holdefer_dynamic_2009-1}),
creating a wide
dynamic range. Modeling and experimental results described in 
\cite{de_schutter_simulated_1994,de_schutter_using_1999},
however, seem to show that the frequencies of Purkinje cells are intrinsically
not so irregular, and that despite their complicated dendritic tree the summation
of inputs can happen independently of their location and distribution. Indeed,
the irregularity in simple spike inter-spike intervals seems to arise because of MLI inputs.
Moreover, in the dendritic tree there exist voltage-gated calcium channels that
amplifiy distant focal inputs more than proximal ones, canceling cable 
attenuation and making the somatic response largely independent of input
location.

Further considerations regarding the general suitability of the PRC 
representation are presented in \cite{smeal_phase-response_2010}. 
For the specific case of
Purkinje cells, the main concerns when using the theory of stochastic
synchronization may be the variability in firing rates, and the fact that
the PRCs change depending on this firing rate. These issues are not fully
resolved, but possible answers are outlined in the last subsection.

\subsection{Several types of synchrony}

Experimental measurements of simple-spike synchrony in Purkinje cells
may point to separate mechanisms, which could depend on the distance
and alignment of the cells, or on whether they depend on mossy fiber
inputs.

Synchrony between PCs less than 150 micrometers apart is perhaps the most
commonly found. Some of its characteristics are 
being present in spontaneous 
\cite{bell_discharge_1969,ebner_temporal_1981,de_solages_high-frequency_2008}
and evoked activity \cite{ebner_temporal_1981},
appearing in cross-correlograms as a sharp peak near
zero lag on top of a peak with long tails 
\cite{shin_dynamic_2006,de_solages_high-frequency_2008}, or
sometimes with two peaks near $\pm$5 ms \cite{de_solages_high-frequency_2008}.
The peaks seem to arise from alignment
of the spikes delimiting pauses in simple spikes, 
and the long tails by rate modulation \cite{shin_dynamic_2006}.

On the other hand, synchrony between PCs several hundred micrometers apart
often happens between cells aligned in the parallel fiber direction 
\cite{ebner_temporal_1981,heck_-beam_2007,bosman_encoding_2010}.
The results in \cite{wise_mechanisms_2010} showed more frequent synchronization
along the rostrocaudal direction, possibly for cells in the same
functional module. Interestingly, cells with complex spike synchrony were more 
likely to have synchronization of simple spikes, and there were
aligned pauses (with mean duration of 129 ms, longer than the pauses
of ~20 ms in \cite{shin_dynamic_2006}).
The results in \cite{bosman_encoding_2010} have some characteristics in common 
with \cite{wise_mechanisms_2010}, with 
complex spike synchronization happening orthogonally to parallel fiber beams
in cells that tended to respond to the same stimuli with distances around
230 micrometers. Thus, synchrony may be found between distant PCs aligned
orthogonally to the parallel fiber axis when those cells have similar response
properties.

Other characteristics of the synchrony between distant PCs is that it may be 
observed in spontaneous activity, or may arise from
inputs, and in this latter case the synchrony may not be explained by
firing rate modulation 
\cite{ebner_temporal_1981,heck_-beam_2007,wise_mechanisms_2010}.
The peak in the cross-correlogram may last for long periods,
especially when afferent inputs are present 
\cite{ebner_temporal_1981,heck_-beam_2007}.

These data suggest that synchrony may have more than one
underlying cause. Broad peaks in  cross-correlograms among nearby PCs may
come from comodulation; alignment of pauses could arise from a combination
of common ascending axon inputs and recurrent inhibition. Synchrony between 
distant PCs along the parallel fiber axis presumably arises from
parallel fiber interactions. Synchrony among distant PCs not aligned in the
parallel fiber axis could possibly arise due to common input, but recurrent
inhibition can't be totally dismissed, although coherence in simple spike
response decays with distance \cite{de_solages_high-frequency_2008}, 
making this less likely. The
fact that for distant PCs simple spike synchronization is more likely when
there is complex spike synchronization \cite{wise_mechanisms_2010} suggests a 
learning mechanism may be in place.

\subsection{Cerebellar models, firing rates, and everything else}

The mechanism of stochastic synchrony does not require to assume any
particular theory of cerebellar function, it merely expands the
space of plausible mechanisms to implement them.
There are several ways in which stochastic synchrony could be
an important factor for cerebellar function, and they all depend on the
physiological details. What is assumed about those details usually
depends on what is assumed about the function of cerebellar
microcircuitry.
In this section I present some examples of synchronization playing a role
in different hypotheses of cerebellar function.

We start by working under the assumption that the cerebellum is
mainly involved in the control of sensory data acquisition 
\cite{bower_is_1997}.
Under this assumption, parallel fibers perform a modulatory role, and
most of the evoked simple spike activity depends on ascending axons, which 
produce synchronous stimulation capable of making the membrane voltage
cross threshold \cite{bower_organization_2002,santamaria_modulatory_2002}. 
Synchrony among nearby Purkinje cells can happen
naturally because of the common ascending axon activation. Under these
assumptions, a
suggestion is that the role of parallel fibers could be to regulate
the amount of synchrony that happens among the cells activated by
ascending axons, and in this way allow the sensory and motor context
to affect the cerebellar response. Moreover, plasticity mechanisms 
elicited by complex spikes could reshape the way in which these
contexts affect synchrony.

In the context of the hypothesis that the cerebellum controls the 
acquisition of sensory information, it may be assumed that the
CF-mediated plasticity acts as a homeostatic mechanism that
regulates the firing rate of Purkinje cells through a negative 
feedback loop \cite{de_schutter_cerebellar_1995,bengtsson_cerebellar_2006}. 
A variation of this 
would be to assume that the CF-mediated plasticity also acts to
equalize the simple spike firing rates of Purkinje cells, which
would permit a higher degree of stochastic synchronization.
A second variation could come if we assume that complex spikes
serve not only a homeostatic function, but constitute a learning 
mechanism by firing in response to sensory stimuli which tend to
require a response, such as an air puff to the eye, the stretching
of a tendon, or a drop in blood pressure. In this second variation, 
the GC activity that tends to happen in
in the absence of complex spikes in a given PC will produce strong
excitatory and inhibitory synapses, whereas GC activity that
tends to happen in conjunction with simple spikes will produce 
weak synapses. The strong, highly correlated input from the
ascending axons will entrain nearby Purkinje cells as long as 
their subthreshold potentials remain similar, since it has a purely
excitatory effect (a type I PRC). Strong and
heterogeneous PF input, like the one from GC activity unrelated
to complex spikes has the ability to desynchronize the
phases of the subthreshold potentials, whereas the weak synapses
of PF inputs correlated with complex spikes cannot disrupt the
synchronization.

Indeed, when sensory evoked granule cell inputs happen in the absence of
complex spikes LTP takes place in the PF-PC synapse 
\cite{lev-ram_new_2002,coesmans_bidirectional_2004,wang_long-term_2009,
schonewille_purkinje_2010}.
At the PF-MLI synapse, inputs that are not
paired with CF activity can experience LTP or LTD in stellate cells,
\cite{rancillac_synapses_2004}, and STD in basket cells 
\cite{bao_target-dependent_2010}. Moreover, PF stimulation
that is unpaired with CF activity increases the size of PC receptive
fields, and the size of MLI receptive fields is decreased 
\cite{jorntell_reciprocal_2002}.
Finally, when there is CF activity that does not happen in conjunction
with complex spikes there is an induction of LTP in the MLI-PC synapses,
known as rebound potentiation \cite{kano_synaptic_1992}. 
These experiments suggest that
the GC activity that is generally not associated with complex
spike activity in a PC has strong excitatory and inhibitory synapses,
large excitatory receptive fields, and a variety of responses in the
PF-MLI synapse.

The opposite scenario seems to happen when GC inputs happen in
conjunction with CF inputs. In this case there is LTD in both
excitatory 
\cite{ito_climbing_1982,ito_synaptic_1993,mittmann_linking_2007,
hirano_long-term_2013} and inhibitory
\cite{mittmann_linking_2007} synapses. In addition, the size of MLI receptive
fields increases, while the size of PC receptive fields 
decreases \cite{jorntell_reciprocal_2002}. The GC activity that is associated with
complex spikes will thus tend to have weaker synapses at
the Purkinje cell, and perhaps broader inhibition. If the PF
activity encodes a sensory context, and if complex spikes arise
in response to sensory stimuli that tend to require a response,
then the contexts associated with the need for a response would
be the ones that disrupt synchrony the least. An increase in
synchrony for particular sensory contexts could complement effects
on the firing rate of Purkinje cells, creating an output signal
from cerebellar cortex that can be used both for behavior 
generation and as a training signal for downstream plasticity.
This creates a hybrid model of cerebellar function, where there
could could be control of sensory information, but this could be
modulated by the particular context. In fact, there seems to be
no strong {\it a priori} reason to assume that the cerebellum 
couldn't perform both selection of relevant sensory information 
and correction of behavior execution.

A rather different hypothesis about cerebellar function comes
from the view of the cerebellum as an adaptive filter (e.g.
\cite{fujita_adaptive_1982,thach_what_1998,dean_cerebellar_2010}).
This view could be said to encompass
the models based on the classical Marr-Albus framework
\cite{marr_theory_1969,albus_theory_1971}. Typically, under this view the parallel
fibers can drive the firing rate of Purkinje cells, and complex
spikes signal peformance errors. The PF-PC synapses that are
positively correlated with errors will undergo LTD, whereas
the PF-PC signals that are negatively correlated with errors
will undergo LTP, so that behavior execution is driven by the
signals that minimize error. A more complete version of this hypothesis
could consider the changes in synchronization that come from
CF-mediated plasticity, as exemplified below.

One view of how CF-mediated plasticity could work in the 
Marr-Albus framework is that this plasticity increases the 
synchrony of the cells not correlated with complex spike activity,
because of the same basic plasticity mechanisms outlined above
(notice that in the previous hypothesis these mechanisms would
lead to desynchronization rather than synchronization). We could
assume that GC activity not associated with complex spikes 
produces strong excitatory and inhibitory synapses that grow until
saturation, and also adopt the key assumption that when synapses 
grow until saturation they become homogeneous, with an equivalent 
PRC of type II. This would lead to a larger degree of synchrony from
signals negatively correlated with complex spikes, complementing
the effect of conjunctive LTD in the PF-PC synapse. On the other hand,
the weak heterogeneous inputs from signals associated with complex
spikes would lead to a low degree of synchronization. The dynamics of
this system could correspond to a generalization of the results 
in \cite{ly_synchronization_2009}. In addition, the silencing of synapses 
from conjunctive
LTD could lead to a decrease of input correlation for particular
patterns of activity, as mentioned in the results section. This variation
of the basic Marr-Albus-Ito hypothesis could explain to some extent
the experimental results which question the sufficiency of conjunctive
LTD for cerebellar learning (summarized in the introduction of this 
paper). For example, the deep hyperpolarizations produced by synchronous
simple spikes could mediate plasticity in the targets of Purkinje cells,
which would explain why the inhibition of Purkinje cells seems to be
necessary for consolidation of motor learning \cite{wulff_synaptic_2009}.

To finish this paper, I present a general hypothesis about stochastic
synchrony that is agnostic to the specific function performed by
the cerebellum, but leads to different predictions to the hypotheses
described above. As before, the basic idea is that stochastic
synchrony is modulated in various functional modules of cerebellar
cortex according to an afferent/efferent context provided by 
parallel fibers from other modules.
An outline of how this could happen is as follows:
First, afferent/efferent input increases the firing rate of
Purkinje cells in one or more parasagittal cerebellar modules 
due to granule cell ascending axons. The firing rate modulation
has a triple effect: it changes the equivalent PRC shapes for the PF 
inputs to type II, it increases the amplitude of those PRCs, and
it equalizes the firing rates of a subset of Purkinje cells.
These effects propitiate stochastic synchrony among the cells activated
with similar firing rates that receive common parallel fiber input.
The synchrony acts as a gain mechanism on the response of the DCN
cells receiveing converging inputs from the responding Purkinje cells.

The reason why the input could change the PRC shape to type II is related 
to how the PRC is modulated by the cell's firing rate. Measurement of
excitatory PRCs in the Purkinje cell indicate that this curve is
flat at low firing rates, but at higher firing rates it presents
a peak at later phases \cite{phoka_new_2010}. It is not known how the 
inhibitory
inputs change their PRC according to the firing rate, but if this
PRC remains flat or develops a peak for early phases then the increase
in firing rate could generate a type II PRC. 

The reason why the activity produced by the ascending axons could increase 
the amplitude of the PRCs is because of the active channels present in the 
PC dendrites, combined with the fact that the synapses from the ascending
axons would increase the amount of intracellular calcium. Similarly, the 
activity induced by the ascending axons could result in subsets of
Purkinje cells with similar firing rates, which could depend on
a gain modulation produced by complex spikes. Moreover, complex spikes
may elicit temporary alignments of simple spike firing rates in
different cells. Following a complex spike a Purkinje cell tend to
go through a pause, facilitation, and suppression, and the simple
spike frequency tends to be reduced \cite{de_zeeuw_spatiotemporal_2011}.

The role of complex spikes under this general hypothesis of stochastic
synchrony could be merely a homeostatic one, the equalization of
firing rates, selection of error-reducing inputs, or the selection
of inputs that predict errors or sensory events. There are assumptions
which could make any of these a viable candidate, and figuring out
the right ones (as well as determining whether stochastic synchrony
plays a role at all in cerebellar function) will require further
experimental and modeling work.



\section*{Acknowledgements}
This work was supported by the ARL/GDRS RCTA project under Cooperative 
Agreement Number W911NF-10-2-0016.

\noindent The author wishes to thank RC O'Reilly and GB Ermentrout for helpful
comments and suggestions.

\bibliographystyle{bmc-mathphys} 
\bibliography{library}      

\end{document}